\documentclass[aps,prb,showpacs,twocolumn,groupedaddress,floatfix]{revtex4-1}

\usepackage{amsmath}
\usepackage{amssymb}
\usepackage{graphicx}
\usepackage[english]{babel}
\usepackage{blindtext}
\usepackage[utf8]{inputenc}

\newcommand{\AlTwoOThree}{Al$_{\mathrm{2}}$O$_{\mathrm{3}}$}
\newcommand{\MgO}{MgO}
\newcommand{\SiOTwo}{SiO$_{\mathrm{2}}$}

\begin{document} 
\title{Microscopic theory of electron absorption by plasma-facing surfaces}

\author{F. X. Bronold and H. Fehske} 
\affiliation{Institut f{\"ur} Physik,
             Ernst-Moritz-Arndt-Universit{\"a}t Greifswald,
             17489 Greifswald,
             Germany}

\date{\today}
\begin{abstract}
We describe a method for calculating the probability with which the wall of 
a plasma absorbs an electron at low energy. The method, based on an invariant 
embedding principle, expresses the electron absorption probability as the 
probability for transmission through the wall's long-range surface potential 
times the probability to stay inside the wall despite of internal backscattering. 
To illustrate the approach we apply it to a \SiOTwo\ surface. Besides emission of 
optical phonons inside the wall we take elastic scattering at imperfections of 
the plasma-wall interface into account and obtain absorption probabilities 
significantly less than unity in accordance with available electron-beam scattering 
data but in disagreement with the widely used perfect absorber model.
\end{abstract}
\maketitle

\section{Introduction} 

A surface facing a plasma collects electrons from the plasma more efficiently 
than it looses electrons due to neutralization of ions and/or 
de-excitation of radicals. It acquires thus a negative charge triggering in 
turn an electron-depletion layer in front of it--the plasma sheath--shielding 
the plasma from the surface. Although known since the beginning of modern plasma 
physics~\cite{MSL26} a quantitative understanding of electron accumulation 
by plasma walls is still lacking~\cite{Franklin76}. This is only due partly 
to unresolved materials science aspects, such as, chemical contamination 
and/or mechanical destruction of the surface by the plasma. It is also 
because little is known fundamentally about the interaction of electrons with 
surfaces at the energies relevant for plasma applications.

Electrons interacting with solid surfaces in the divertor region
of fusion plasmas~\cite{Tolias14a}, dielectric barrier discharges~\cite{BWS12,TBW14,PS15},
dusty plasmas~\cite{RBP10,Ishihara07,FIK05}, Hall thrusters~\cite{DRF03,BMP03}, 
or electric probe measurements~\cite{LMB07} have typically energies below 
10\,eV, much less than the electron energy used in surface 
analysis~\cite{Cazaux12,SW13,Werner01} or materials processing~\cite{KBK14}. The
energies there are a few 100\,eV, an energy range, where the physical 
processes involved, backscattering and secondary electron emission, are sufficiently 
well understood~\cite{GT03,SF02,DJT00,Vicanek99,DRW95,Tofterup85,KO72,Dashen64} to 
make these techniques reliable tools of applied science. Much less is however 
known about these processes below 100\,eV and hence in the energy range relevant
for plasmas. In particular, the backscattering probability of a low-energy electron, 
and closely related to it, the probability with which it is absorbed  
is basically unknown. 

Although electron absorption (sticking) and backscattering are important processes
for bounded plasmas there is no systematic effort to determine their probabilities
either experimentally or theoretically. The electron sticking probability, for instance, 
is usually assumed to be close to unity~\cite{SEK13,SAB07,KDK04,BMG05,USB00}, 
irrespective of the energy and angle of incident or the wall material (perfect 
absorber assumption~\cite{Alpert65}). The need to overcome this assumption has 
been strongly emphasized by Mendis~\cite{Mendis02} but the model calculations
he refers to are based on classical considerations not applicable to electrons. 

In a recent work~\cite{BF15} we proposed therefore a quantum-mechanical 
approach for calculating the electron sticking probability. The method is based
on two important facts noticed by Cazaux~\cite{Cazaux12}: (i) low-energy electrons 
do not see the strongly varying short-range potentials of the surface's ion cores 
but a slowly varying surface potential and (ii) they penetrate deeply into the 
surface. For \AlTwoOThree, for instance, the average electron penetration depth at 
a few eV is around 200\AA~\cite{Hickmott65}. The sticking probability for an 
electron approaching the wall of a plasma can thus be expressed by the transmission
probability for the long-ranged surface potential times the probability to remain 
inside the wall despite of internal backscattering. Essential for our approach is 
the invariant embedding principle~\cite{Dashen64,Vicanek99,GT03,GP07}. It allows us 
to extract from the overwhelming number of electron trajectories the few backwardly 
directed ones most relevant for sticking.
So far we applied the method to \MgO~\cite{BF15} obtaining excellent 
agreement with electron-beam scattering data~\cite{CF62}. In this work we consider 
\SiOTwo\ finding again good agreement with beam 
data~\cite{Dionne75}. In both cases the sticking probability is 
energy- and angle-dependent as well as significantly less than unity.

The remaining part of the paper is organized as follows.
In Section \ref{Formalism} we describe our microscopic approach for calculating 
electron absorption and backscattering probabilities in more detail than 
previously~\cite{BF15}, focusing in particular on the invariant embedding principle 
and its linearization making the approach numerically very efficient. Section 
\ref{Results} presents results for \SiOTwo, an in-depth discussion of the model we 
proposed for the description of imperfect plasma-wall interfaces, and a calculation of 
orbital-motion limited grain charges beyond the perfect absorber model for electrons.
Concluding remarks are given in Section \ref{Conclusions}.

\section{Formalism}
\label{Formalism}

The method we developed for calculating the probability with which a low-energy 
electron is absorbed by a surface is general~\cite{BF15}. It can be applied to metallic 
as well as dielectric surfaces. To be specific we consider in this work a
dielectric \SiOTwo\ surface as an example. 

\begin{figure}[t]
\begin{minipage}{0.68\linewidth}
\includegraphics[width=\linewidth]{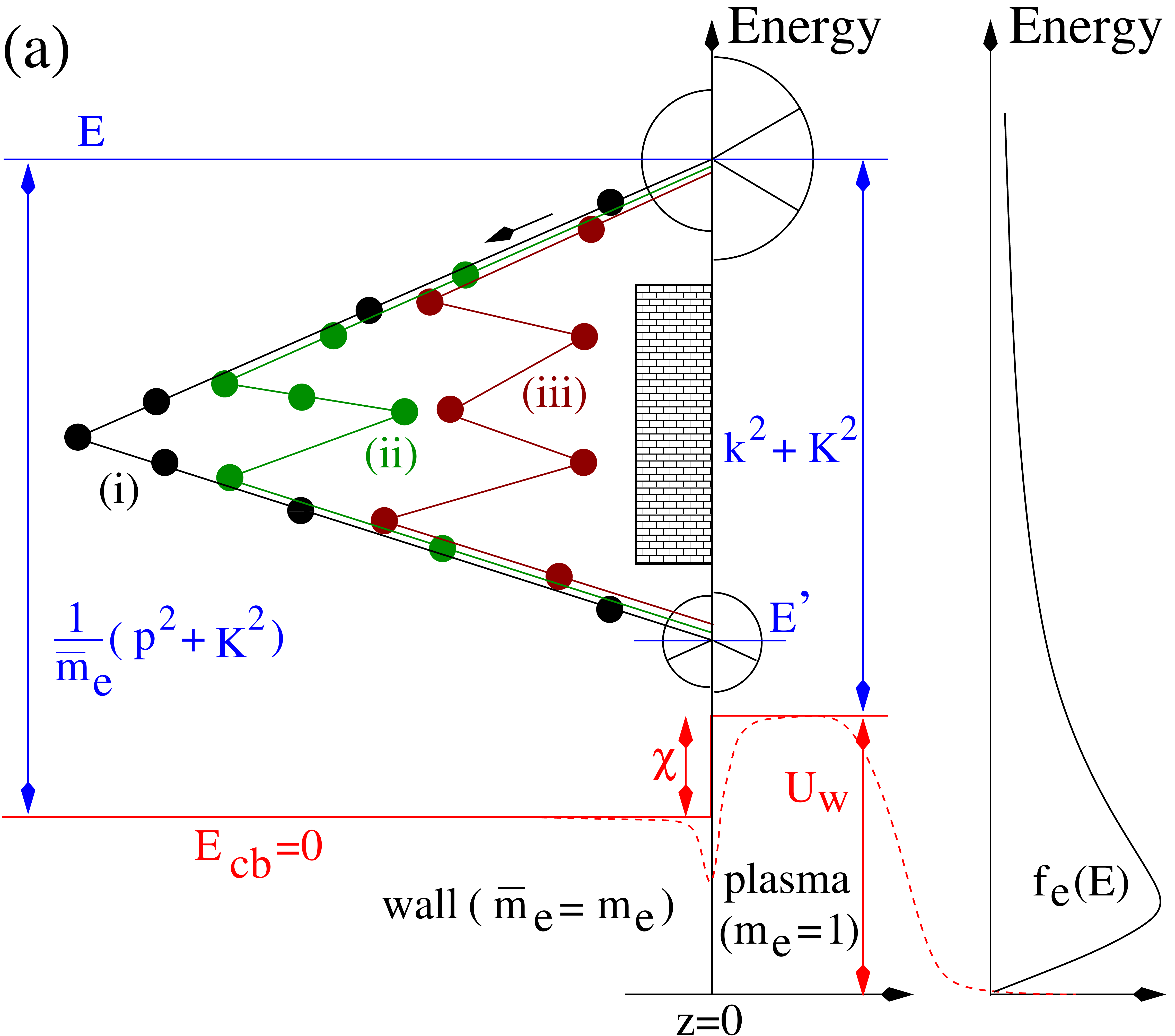}
\end{minipage}\begin{minipage}{0.32\linewidth}
\includegraphics[width=0.99\linewidth]{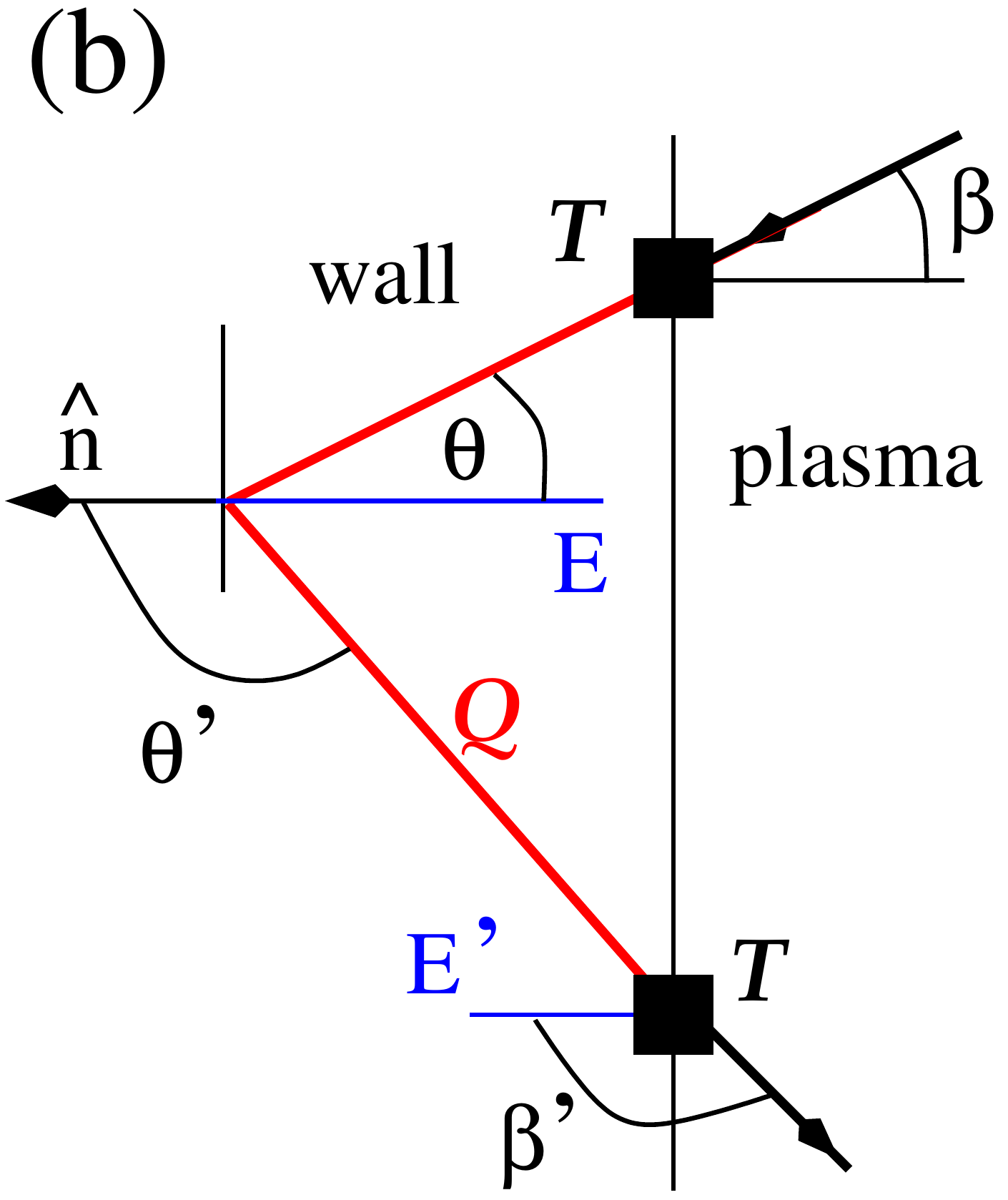}
\end{minipage}
\caption{(a) (color online) Interface model used in our calculation. 
The surface potential (dashed red line) is approximated by a potential step of 
height $\chi$ (solid red line). Three scattering trajectories (i)--(iii) due
to emission of optical phonons inside the wall symbolized by bullets are shown 
each having the same number of total but a different number of backscattering events 
and bringing an electron entering the wall at energy $E$ back to the plasma at an 
energy $E^\prime$. The half circles denote the moduli of the electron momenta inside
and outside the wall. Also shown is the energy distribution $f_e(E)$ the approaching
electron may have. (b) Illustration of Eqs.~\eqref{StickCoeff}--\eqref{EscapeProb}.
The potential step leads to a quantum-mechanical transmission probability ${\cal T}$
whereas the emission of phonons yields a quantity $Q$ to be obtained from the
invariant embedding principle shown in Fig.~\ref{Embedding}. The pre- and
post-collision angles inside $(\theta, \theta^\prime)$ and outside
$(\beta, \beta^\prime)$ the wall are measured with respect to the surface
normal $\hat{n}$.
}
\label{CartoonEPS}
\end{figure}

For a dielectric wall with $\chi>0$ (positive electron affinity) such as \SiOTwo\
the potential energy of an electron across the plasma-wall interface has roughly the
form~\cite{HBF12} shown, together with other aspects of our approach, in Fig.~\ref{CartoonEPS}a. 
An electron approaching the interface from the plasma has to overcome the wall 
potential $U_w$. Once it is inside the wall it occupies the conduction band and sees thus 
a potential barrier $\chi$. Since it is the kinetic energy of the electron in the 
immediate vicinity of the interface which determines sticking and backscattering 
probabilities, while the variation of the wall potential $U_w$ is on the scale of the 
Debye screening length, much larger than the scale on which the surface potential 
varies, the relevant part of the electron potential energy is essentially a three-dimensional 
potential step with height $\chi$ and electron mass mismatch $\overline{m}_e=m_e^*/m_e < 1$, 
where $m_e^*$ is the effective electron mass in the conduction band of the wall and $m_e$ is 
the bare electron mass, as illustrated by the solid red line in Fig.~\ref{CartoonEPS}a.

The potential step gives rise to quantum-mechanical reflection and transmission. For
the situation shown in Fig.~\ref{CartoonEPS}a, that is, a wall (plasma) occupying the 
$z<0$ ($z>0$) half space and an energy scale for which $E_{\rm cb}=U_w-\chi\equiv 0$, 
the transmission probability for an electron coming from the plasma, and having 
thus a kinetic energy $E-\chi>0$, is given by~\cite{WY79}
\begin{align}
{\cal T}(E,\xi)=\frac{4\overline{m}_e k p}{(\overline{m}_e k + p)^2}
\label{Trans}
\end{align}
with $k=\sqrt{E-\chi}\,\xi$ and $p=\sqrt{\overline{m}_e E}\,\eta$ the $z-$components 
of the electron momenta outside and inside the wall. In \eqref{Trans}
and the formulae below we measure length in Bohr radii, energy in Rydbergs, and mass 
in electron masses implying inside the wall the electron mass is simply the mass 
mismatch $\overline{m}_e$. The signs of
$k$ and $p$ in~\eqref{Trans} are always the same. We can thus define the direction
cosines $\xi$ and $\eta$ referenced, respectively, to the electron momenta outside and 
inside the wall, by their absolute values: $\xi=|\cos\beta|$ and $\eta=|\cos\theta|$ 
(see Fig.~\ref{CartoonEPS}b for the definition of the angles). This choice is also
convenient for the theoretical description of internal backscattering which we address
later. Since the potential varies only perpendicularly to the surface the lateral 
momentum $\vec{K}$ is conserved. Together with the conservation of energy, 
$E=\chi+k^2+\vec{K}^2=(p^2+\vec{K}^2)/\overline{m}_e$, this leads to 
\begin{align}
1-\eta^2 = \frac{E-\chi}{\overline{m}_e E}\big(1-\xi^2\big)~
\label{etaxi}
\end{align}
connecting the direction cosines $\eta$ and $\xi$. From \eqref{etaxi} follows
that an electron approaching the wall with kinetic energy $E-\chi=\vec{K}^2 + k^2>0$ 
enters it only when $\xi$ is larger than
\begin{align}
\xi_c=\left\{\begin{array}{ll}
0 & ~{\rm for}~~ \chi < E < E_0 \\
\sqrt{1-\frac{\overline{m}_e E}{E-\chi}} & ~{\rm for}~~ E>E_0
\end{array}\right. 
\label{xic}
\end{align}
with $E_0=\chi/(1-\overline{m}_e)$. For $\xi$ less than $\xi_c$ the electron is in 
an evanescent wave with $p^2<0$ and thus totally reflected~\cite{GB89}. In addition, 
the requirement~\eqref{etaxi} may instantaneously reduce the electron's perpendicular 
kinetic energy to less then the electron affinity $\chi$ once it crossed the 
interface from the plasma side, that is, $p^2/\overline{m}_e  < \chi$ even without 
inelastic scattering. For mass mismatch $\overline{m}_e < 1$, applicable to
\SiOTwo, \MgO, \AlTwoOThree, this happens when 
$\xi<\sqrt{1-\overline{m}_e}$. Provided the electron cannot gain energy by inelastic 
scattering, as it is the case for dielectric walls at room temperature, it will have 
no chance to ever come back to the plasma. 

The transmission probability ${\cal T}(E,\xi)$ is not identical with the sticking 
probability. It captures only the ballistic aspect of electron absorption by the wall
and is at best an upper bound to it. 
Once the electron is inside the wall it suffers elastic as well as inelastic scattering. Both 
may push the electron back to the interface and, after successfully traversing the surface 
potential in the reverse direction, eventually back to the plasma. Hence, we expect 
the sticking probability $S(E,\xi)\le{\cal T}(E,\xi)$. To take scattering inside the wall 
into account we defined $S(E,\xi)$ as the probability of an electron hitting the wall 
from the plasma with energy $E$ and direction cosine $\xi$ not to return to it after entering 
the wall and suffering backscattering~\cite{BF15},
\begin{align}
S(E,\xi) = {\cal T}(E,\xi)[1-{\cal E}(E,\xi)]~,
\label{StickCoeff}
\end{align}
where  
\begin{align}
{\cal E}(E,\xi)&= 
\int_{\eta_{\rm min}}^1 \!\!\!\!\!\! d\eta^\prime \int_{E^\prime_{\rm min}}^E \!\!\!\!dE^\prime
\rho(E^\prime){\cal B}(E\eta(\xi)|E^\prime\eta^\prime){\cal T}(E^\prime,\xi(\eta^\prime))
\label{EscapeProb}
\end{align}
is the conditional probability for the electron to escape from the wall after at least one 
backscattering event. The lower integration limits, $\eta_{\rm min}=\sqrt{\chi/E}$ 
and $E^\prime_{\rm min}=\chi/\eta^{\prime\, 2}$, ensure that only events are counted 
for which the perpendicular post-collision energy $p^{\prime\, 2}/\overline{m}_e>\chi$, 
$\rho(E)=\sqrt{\overline{m}_e^3E}/2(2\pi)^3$ is the conduction band's density of states, and
\begin{align}
{\cal B}(E\eta|E^\prime\eta^\prime)=
\frac{Q(E\eta|E^\prime\eta^\prime)}
{\int_0^1 d\eta^\prime \int_0^E dE^\prime \rho(E^\prime)Q(E\eta|E^\prime\eta^\prime)}
\label{Bfct}
\end{align}
is the normalized probability $Q(E\eta|E^\prime\eta^\prime)$ for an electron with energy 
$E$ and direction cosine $\eta$ to backscatter after an arbitrary number of internal scattering 
events to a state with energy $E^\prime$ and direction cosine $\eta^\prime$. Since the 
function $Q(E\eta|E^\prime\eta^\prime)$ describes backscattering, 
$0 < \theta \le \pi/2$ and $\pi/2 < \theta^\prime \le \pi$, implying 
$0\le \eta,\eta^\prime < 1$ as $\eta=|\cos\theta|$. The energy integrals in~\eqref{EscapeProb} 
and \eqref{Bfct} anticipate that at room temperature the electron cannot gain energy from a 
dielectric wall and the functions $\eta(\xi)$ and $\xi(\eta)$ are implicitly defined by~\eqref{etaxi}. 

\begin{figure}[t]
\includegraphics[width=\linewidth]{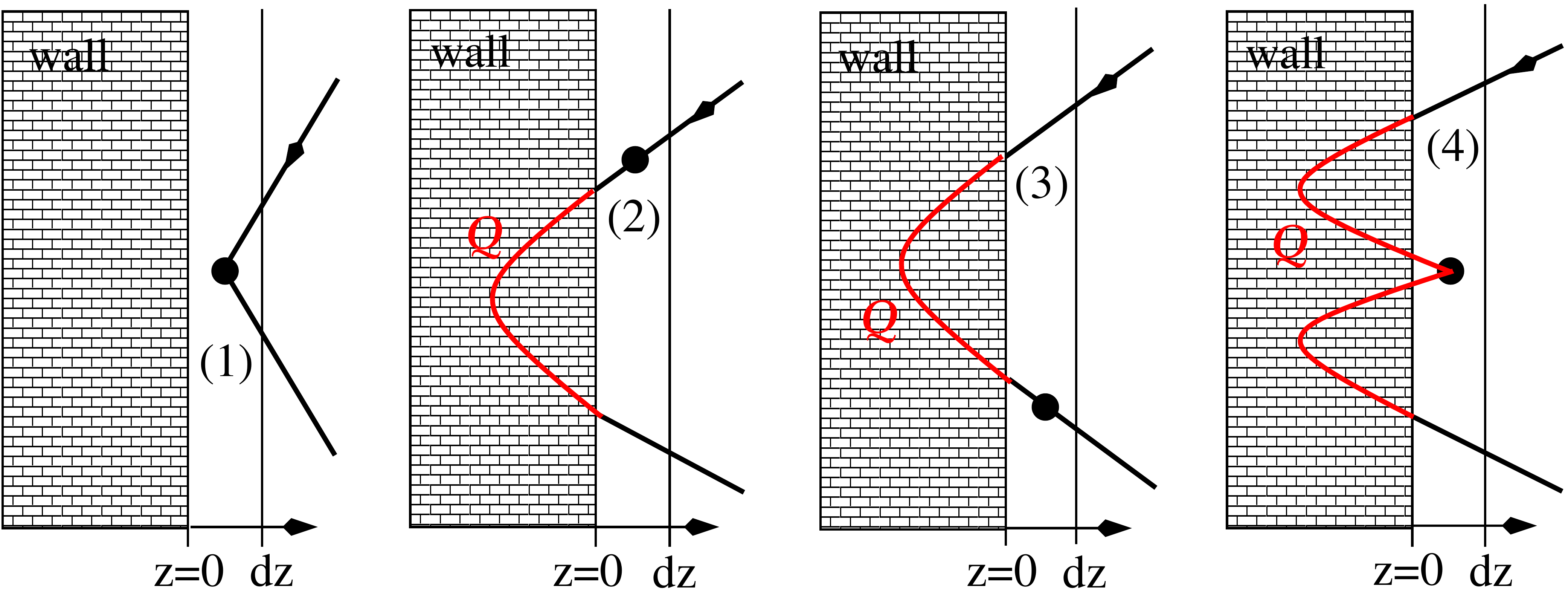}
\caption{(color online) Schematic illustration of the principle of invariant embedding. Due to
the infinitesimally thin additional layer of identical material four additional paths (1)--(4)
may contribute to $Q(E\eta|E^\prime\eta^\prime)$. However, an infinitesimally thin layer of the
same material cannot affect $Q(E\eta|E^\prime\eta^\prime)$. Hence, $Q(E\eta|E^\prime\eta^\prime)$
has to be invariant against the change the paths (1)--(4) induce leading to Eq.~\eqref{NLIntEq}.}
\label{Embedding}
\end{figure}

To obtain the quantity $Q(E\eta|E^\prime\eta^\prime)$ we employ the invariant 
embedding principle~\cite{Dashen64,Vicanek99,GT03,GP07}, the essence of it is shown in 
Fig.~\ref{Embedding}. For our purpose it is extremely powerful since it focuses from 
the start on the backscattering trajectories. Compared to forward scattering  
backscattering is usually much less likely. Constructing thus 
$Q(E\eta|E^\prime\eta^\prime)$, for instance, from the Monte Carlo trajectories mimicking
the solution of the electron's Boltzmann equation, obtained under suitable initial and 
boundary conditions, would be numerically very expensive. 

The principle can be derived as 
follows~\cite{Dashen64}. Imagine to add to the half space filled with wall material an 
infinitesimally thin layer of the same material. As a result the four scattering trajectories 
shown in Fig.~\ref{Embedding} may now additionally contribute to $Q(E\eta|E^\prime\eta^\prime)$. 
However, an infinitesimally thin additional layer of the same material cannot change the 
backscattering properties of the half-space. Hence, $Q(E\eta|E^\prime\eta^\prime)$ has to 
be invariant against this change. 

Defining a convolution for functions depending on the variables $E, \eta, E^\prime$,
and $\eta^\prime$,
\begin{align}
(A\ast B)(E\eta|E^\prime\eta^\prime) &= 
\int_0^\infty \!\!\!dE^{\prime\prime}\int_0^1 \!\!\!d\eta^{\prime\prime}
\rho(E^{\prime\prime})A(E\eta|E^{\prime\prime}\eta^{\prime\prime})\nonumber\\
&\times B(E^{\prime\prime}\eta^{\prime\prime}|E^\prime\eta^\prime)~,
\end{align}
summing up the four paths 
(1)--(4) shown in Fig.~\ref{Embedding} (without the transmission/reflection due to the surface 
potential), and enforcing them not to change $Q(E\eta|E^\prime\eta^\prime)$ yields 
the nonlinear integral equation~\cite{Dashen64,GP07} 
\begin{align}
&\big[\frac{\Pi(E)}{\eta}+\frac{\Pi(E^\prime)}{\eta^\prime}\big]Q(E\eta|E^\prime\eta^\prime)=
G^-(E\eta|E^\prime\eta^\prime) \nonumber\\
&+ (G^+ \!\!\ast Q)(E\eta|E^\prime\eta^\prime) + (Q\ast G^+)(E\eta|E^\prime\eta^\prime) \nonumber\\
&+ (Q\ast G^- \!\!\ast Q)(E\eta|E^\prime\eta^\prime)~,
\label{NLIntEq}
\end{align}
where the kernels 
\begin{align}
G^\pm(E\eta|E^\prime\eta^\prime)=\frac{\delta(E-E^\prime-\omega)}{\eta}K^{\pm}(E\eta|E^\prime\eta\prime)~
\end{align}
encode forward ($+$) or backward ($-$) scattering. Physically, the lhs describes the reduction of 
the probability for the electron to follow any one of the old paths, already included in 
$Q(E,\eta|E^\prime,\eta^\prime)$, while the rhs corresponds to the four trajectories shown 
in Fig.~\ref{Embedding}.

The kernels $G^\pm(E\eta|E^\prime\eta^\prime)$ are scattering rates per length. The non-trivial 
parts $K^\pm(E\eta|E^\prime\eta^\prime)$ depend on the scattering process. They can be obtained
from the golden rule scattering rate per time dividing it by the pre-collision velocity. Below
we consider a \SiOTwo\ surface, where emission of optical phonons dominates the scattering,
leading to~\cite{BF15}
\begin{align}
K^\pm(E\eta|E^\prime\eta^\prime) &= \frac{1}{2\rho(E)}
\big[ (E+E^\prime \mp 2\sqrt{E E^\prime} \eta\eta^\prime)^2 \nonumber\\
&- 4 E E^\prime (1-\eta^2)(1-\eta^{\prime 2})\big]^{-1/2}~.
\label{Kkernel}
\end{align}
The function $\Pi(E)$ also entering~\eqref{NLIntEq} describes the rate per length to make 
any collision. For phonon emission it is given by $\Pi(E)={\rm arcosh}(\sqrt{E/\omega})/E$.

\begin{figure}[t]
\begin{minipage}{0.5\linewidth}
\includegraphics[width=\linewidth]{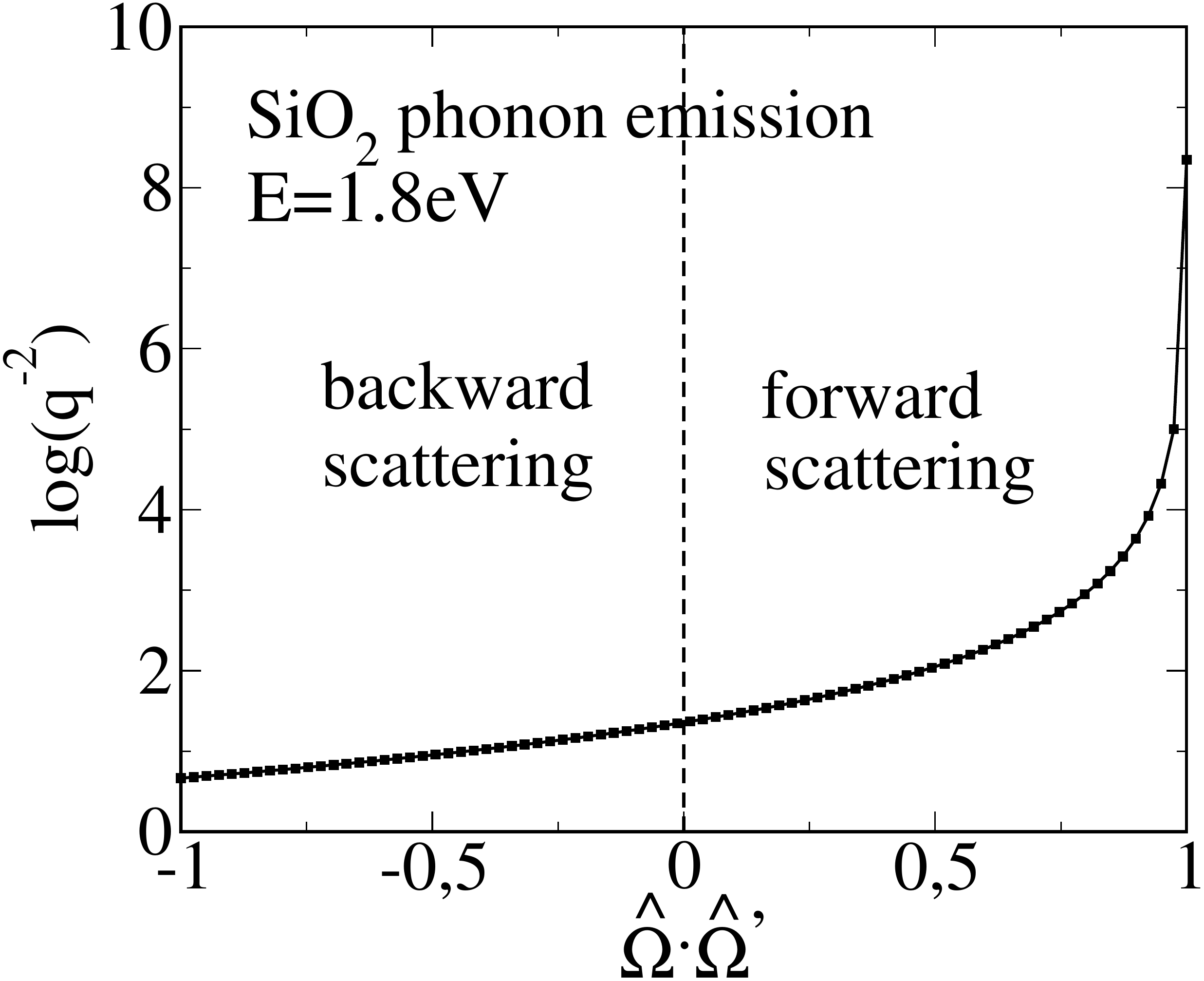}
\end{minipage}\begin{minipage}{0.49\linewidth}
\includegraphics[width=0.85\linewidth]{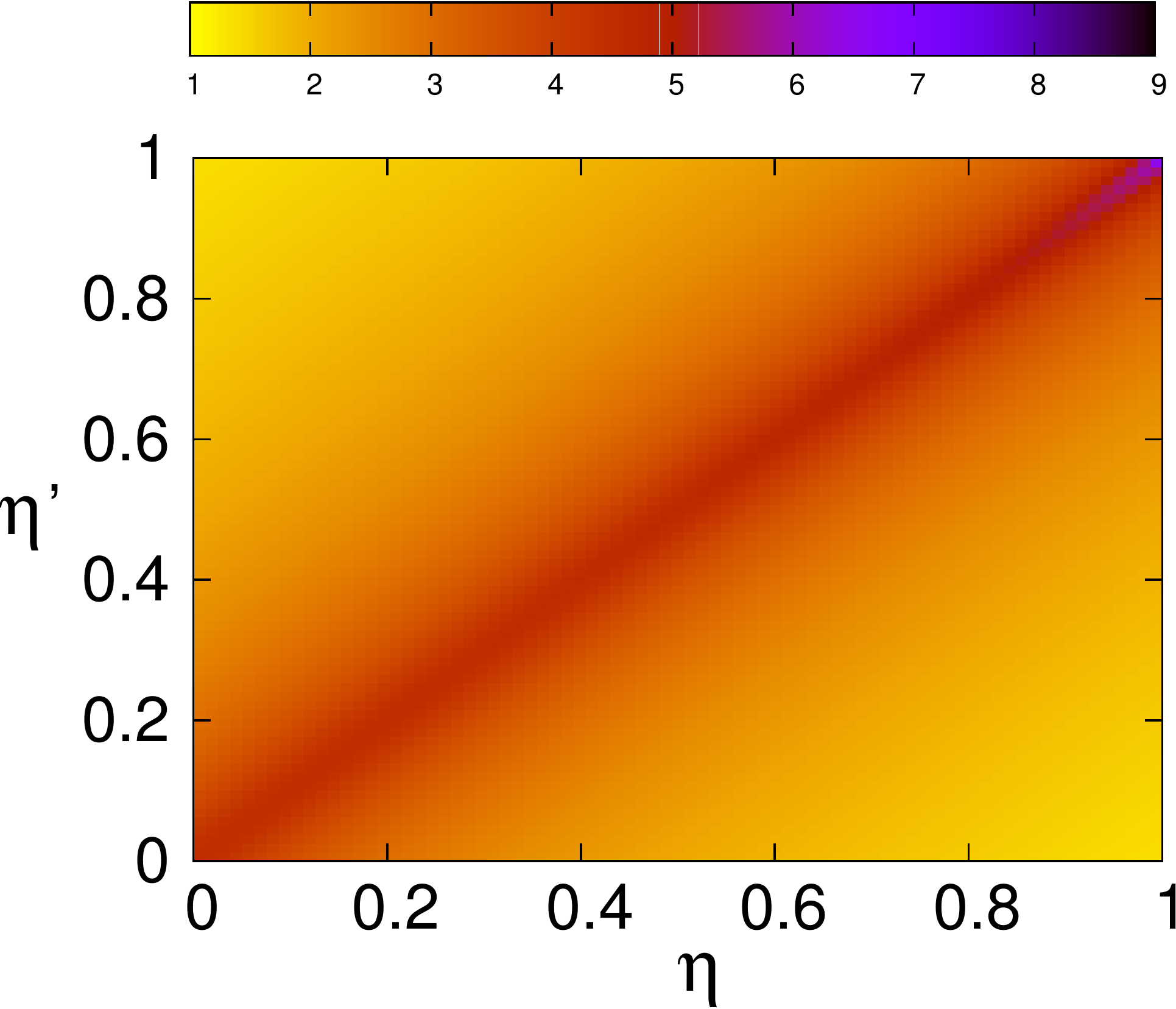}
\end{minipage}
\caption{(color online) On the left is shown for \SiOTwo\ and $E=1.8\,{\rm eV}$ the
square of the inverse of the momentum transfer $q$ as a function of the cosine of the angle 
between the pre- and post-collision momenta. It defines at the low electron densities 
considered in this work the matrix element for phonon emission by the electron entering
the wall. The strong forward peaking 
of the matrix element localizes the forward scattering kernel $K^+(E\eta|E-\omega\eta^\prime)$,
plotted on the right on a log-scale over the whole range of direction cosines $\eta$ and $\eta^\prime$, 
around the diagonal $\eta=\eta^\prime$. A further consequence of the angle dependence of
$q$ is that the backscattering kernel $K^-(E\eta|E-\omega\eta^\prime)$
(not shown) is rather isotropic in the $(\eta,\eta^\prime)$-plane. 
}
\label{XSection}
\end{figure}

In the general case it is hard to work with~\eqref{NLIntEq}. At low energies however
most scattering processes are forwardly peaked. The larger the change of the propagation 
direction the less likely the process. It is thus possible to use the backscattering kernel 
$G^-(E\eta|E^\prime\eta^\prime)$ as an expansion parameter controlling an iterative 
solution of~\eqref{NLIntEq}. Following Glazov and P\'azsit~\cite{GP07} we expand therefore
in a first step the solution of~\eqref{NLIntEq} in the number of the backscattering events. 
In a second step we take advantage of the fact that for dielectric surfaces at room 
temperature scattering arises mainly form the emission of optical phonons with energy 
$\omega$. The electron cannot gain energy by scattering.
It can loose at most the energy it initially had when entering the wall. Expanding thus
$Q(E\eta|E^\prime\eta^\prime)$ also in the number of forward scattering events yields 
a double series which terminates after a finite number of terms. From the differential 
scattering cross section for the materials we are interested in, \SiOTwo, \MgO, and
\AlTwoOThree\, follows moreover that backward scattering due to emission of optical phonons 
is at least two orders of magnitude less likely than forward scattering~\cite{Ridley98}. 
For \SiOTwo\ this can be seen in the left panel of Fig.~\ref{XSection}. Hence, 
writing~\cite{BF15}
\begin{align}
Q(E\eta|E^\prime\eta^\prime)=\sum_{n=1}^\infty\sum_{m=0}^\infty Q^n_{m}(E;\eta|\eta^\prime)
\delta(E-E^\prime-\omega^n_m)
\label{Qexpansion}
\end{align}
with $\omega^n_m=(n+m)\omega$ we can truncate the summation already after a single backward
scattering event, that is, after $n=1$ leading to a linear recursion~\cite{BF15}
\begin{align}
Q^1_{m}(E;\eta|\eta^\prime) &=
F_m(E;\eta|\eta^\prime)Q^1_{m-1}(E-\omega;\eta|\eta^\prime)\nonumber\\
&+Q^1_{m-1}(E;\eta|\eta^\prime)G_m(E-m\omega;\eta|\eta^\prime)~
\label{RR}
\end{align}
for the expansion coefficients with $m=1,...,M_{\rm tot}$, where 
$M_{\rm tot}=\lfloor E/\omega \rfloor -1$ is the number of forward scattering 
events at most possible,   
\begin{align}
F_m(E;\eta|\eta^\prime) &= \frac{K^+(E|E-\omega;\eta)\eta^\prime\rho(E-\omega)}
                        {\eta^\prime \Pi(E) + \eta\Pi(E-(m+1)\omega)}~, \\
G_m(E;\eta|\eta^\prime) &= \frac{\eta\rho(E)K^+(E|E-\omega;\eta^\prime)}
                        {\eta^\prime \Pi(E+m\omega) + \eta\Pi(E-\omega)}~,
\end{align}
and an initialization 
\begin{align}
Q^1_{0}(E;\eta|\eta^\prime) = \frac{\eta^\prime K^-(E\eta|E-\omega\eta^\prime)}
{\eta^\prime \Pi(E) + \eta \Pi(E-\omega)}~.
\end{align}

In deriving the recursion~\eqref{RR} we assumed forward scattering not to change 
the direction cosine at all. This is justified because, as shown in the right panel of 
Fig.~\ref{XSection}, the kernel $K^+(E\eta|E^\prime\eta^\prime)$ is strongly peaked 
for $\eta=\eta^\prime$. The directional change due to forward scattering is thus negligible
and integrals over the direction cosine containing $K^+(E\eta|E^\prime\eta^\prime)$
can be handled by a saddle-point approximation. Forward scattering is then  
encoded in 
\begin{align}
K^+(E|E^\prime;\eta) &=\int_0^1 \!\!\! d\bar{\eta} K^+(E\eta|E^\prime\bar{\eta})
                     =\int_0^1 \!\!\! d\bar{\eta} K^+(E\bar{\eta}|E^\prime\eta)\nonumber\\
                     &=\frac{1}{4\rho(E)\sqrt{EE^\prime}}\log\frac{r(E|E^\prime;\eta)}{q(E|E^\prime;\eta)}~,
\end{align}
where we used the symmetry of $K^+(E\eta|E^\prime\eta^\prime)$ with respect to interchanging 
$\eta$ and $\eta^\prime$ and defined the functions
\begin{align}
r(E|E^\prime;\eta) &= \sqrt{[E+E^\prime]^2 - 4\sqrt{EE^\prime}(E+E^\prime)\eta + 4EE^\prime\eta^2}\nonumber\\
                   & + \sqrt{EE^\prime} - (E+E^\prime)\eta~,\\
q(E|E^\prime;\eta) &= \sqrt{[E-E^\prime]^2 +4EE^\prime\eta^2} - (E+E^\prime)\eta~.
\end{align}
We thus end up with a model similar in spirit to the Oswald, Kasper and 
Gaukler model~\cite{OKG93} for multiple elastic backscattering of electrons from surfaces.

Inserting finally \eqref{Qexpansion} for the backscattering probability into \eqref{EscapeProb} 
and performing the energy integrals yields 
\begin{widetext}
\begin{align}
{\cal E}(E,\xi)=
\frac{\sum_{m=0}^{M_{\rm open}} \int_{\eta_{\rm min}}^1 d\eta^\prime \rho(E-\omega_m^1)
Q_m^1(E;\eta(\xi)|\eta^\prime) {\cal T}(E-\omega_m^1,\xi(\eta^\prime))}
{\sum_{m=0}^{M_{\rm tot}} \int_0^1 d\eta^\prime \rho(E-\omega_m^1)Q_m^1(E;\eta(\xi)|\eta^\prime)}
=\sum_{m=0}^{M_{\rm open}} \int_{\eta_{\rm min}}^1 d\eta^\prime P_m^1(E;\eta(\xi)|\eta^\prime)
\label{Pm1}
\end{align}
\end{widetext}
with $M_{\rm open}=\lfloor (E\eta^2-\chi)/(\eta^2\omega) \rfloor - 1$. Substituting this 
expression for ${\cal E}(E,\xi)$ into Eq.~\eqref{StickCoeff} gives the electron sticking probability 
$S(E,\xi)$ for a clean, homogeneous dielectric wall with positive electron affinity. 

Besides $S(E,\xi)$ the backscattering probability is also of interest
for plasma modeling. Our approach contains two types of backscattering processes: specular 
quantum-mechanical reflection given by ${\cal R}(E,\xi)=1-{\cal T}(E,\xi)$ and diffuse backscattering 
encoded in 
\begin{align}
R_m^1(E;\eta(\xi)|\eta^\prime)={\cal T}(E,\xi)P_m^1(E;\eta(\xi)|\eta^\prime)~ 
\label{Rm1}
\end{align}
with $P_m^1(E;\eta(\xi)|\eta^\prime)$ defined in~\eqref{Pm1}. The latter gives the 
probability for an electron hitting the wall with energy $E$ and direction cosine $\xi$ to end 
up in a state with energy $E^\prime=E-m\omega$ and direction cosine $\xi^\prime=\xi(\eta^\prime)$ 
where $\eta^\prime$ and $\xi^\prime$ are the post-collision direction cosines inside and outside 
the wall. 

\section{Results}
\label{Results}

\begin{figure}[t]
\begin{minipage}{0.49\linewidth}
\includegraphics[width=0.85\linewidth]{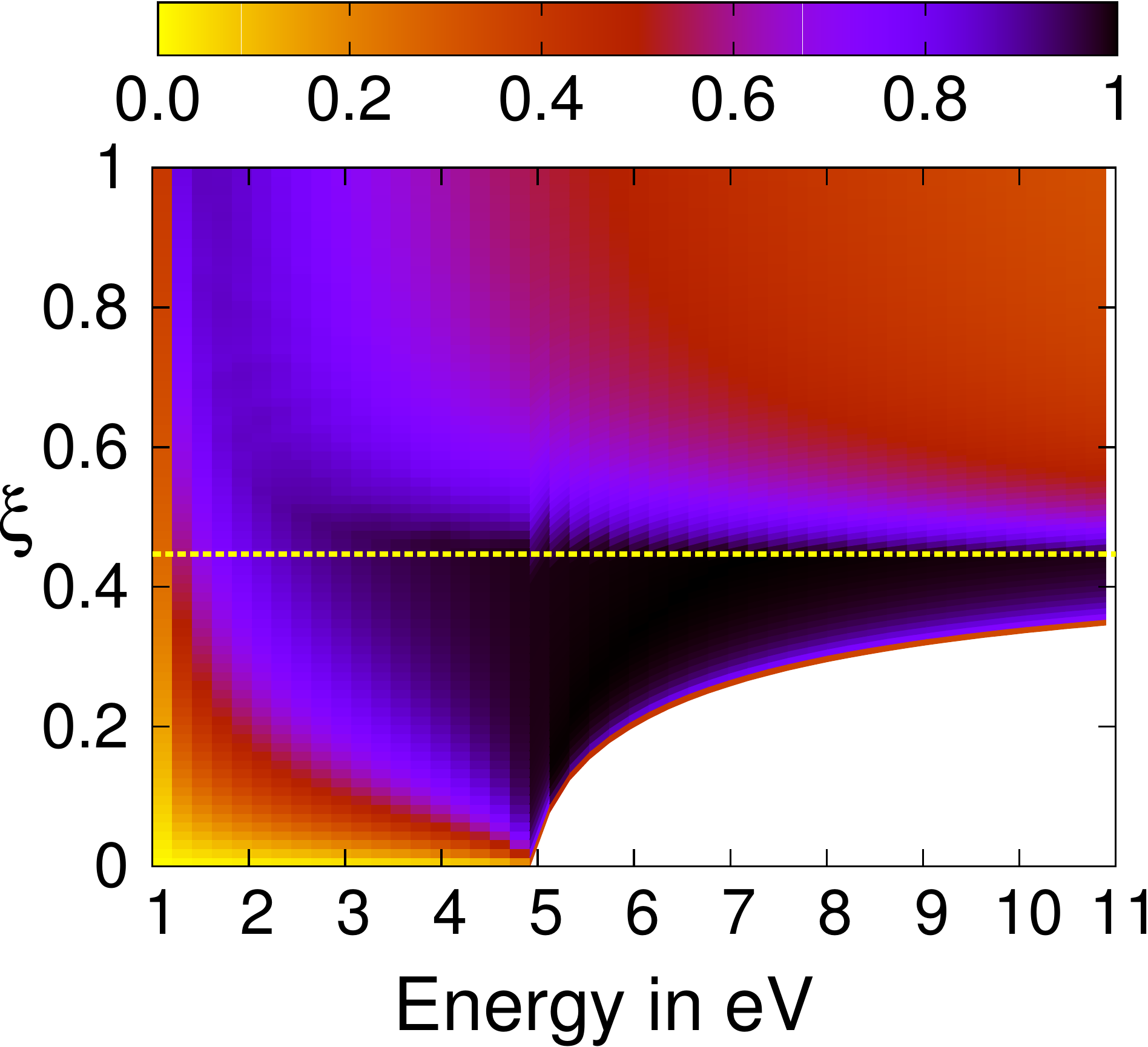}
\end{minipage}\begin{minipage}{0.49\linewidth}
\includegraphics[width=\linewidth]{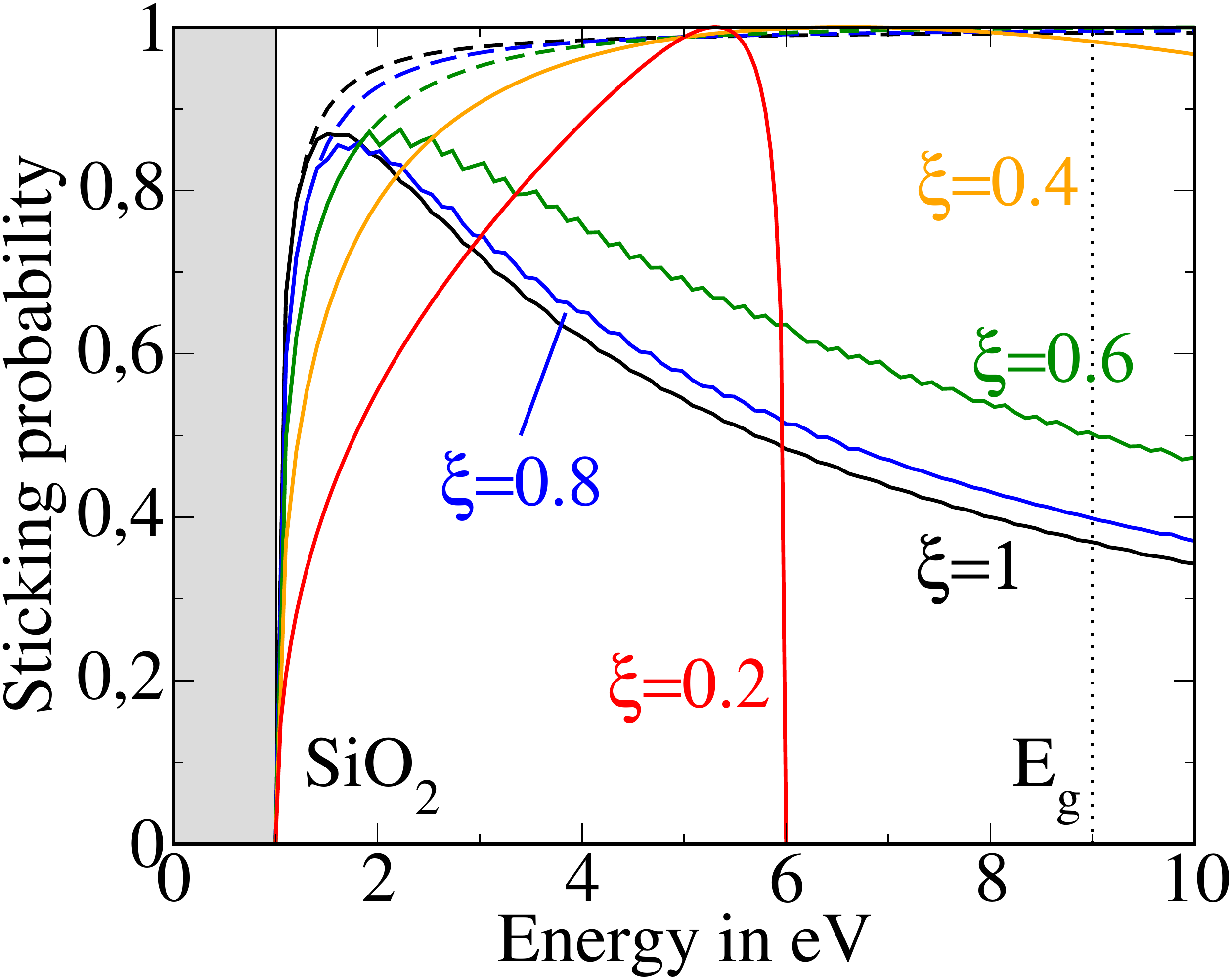}
\end{minipage}
\caption{(color online) Angle-resolved sticking probability $S(E,\xi)$ for a
clean \SiOTwo\ surface. The left panel shows $S(E,\xi)$ for the whole range
of direction cosines $\xi$ and energies $E\le 11\,{\rm eV}$.
Total reflection takes place in the white region. Below the yellow dotted line,
indicating $\xi=\sqrt{1-\overline{m}_e}$, inelastic backscattering has no
effect on the sticking probability. In the right panel $S(E,\xi)$ (solid line)
and ${\cal T}(E,\xi)$ (dashed line) are plotted as a function of $E$ for
representative $\xi$. The grey area denotes the energy range of the conduction
band.
}
\label{AngleResolved}
\end{figure}

We now apply our approach to a \SiOTwo\ surface characterized by 
$\overline{m}_e=0.8$~\cite{SF02}, $\chi=1\,{\rm eV}$~\cite{BBS02}, 
$\omega=0.15\,{\rm eV}$~\cite{SF02}, and $E_g=9\,{\rm eV}$~\cite{VMC11}. 
For an actual \SiOTwo\ surface the parameters may deviate from these values
depending on material science aspects which we do not address in this work.

Numerical results for $S(E,\xi)$ are shown in Fig.~\ref{AngleResolved}. First, we 
focus on the left panel showing data over the whole range of direction cosines $\xi$
and energies $E$ up to $11\,{\rm eV}$. For $E>E_g=9\,{\rm eV}$ our results are only 
rough estimates since an electron entering the wall at these energies can already 
create electron-hole pairs across the band gap. This Coulomb-driven process is not
included. It can be treated 
in the same spirit leads however to a recursion containing energy integrals making 
the numerical treatment more demanding. The white area in the plot 
for $S(E,\xi)$ indicates the region in the $(E,\xi)$-plane where total reflection 
occurs. It is smaller than for \MgO~\cite{BF15} because $\overline{m}_e$ is larger 
for \SiOTwo. Below the dotted yellow line inelastic backscattering due to emission
of phonons is irrelevant for sticking because conservation of lateral momentum 
and total energy force the perpendicular energy of the electron to drop below the 
potential step $\chi$ once it crossed the interface from the plasma side. 
It is hence already confined by quantum-mechanical transmission alone. Only above 
the dotted yellow line inelastic backscattering may bring the electron back 
to the interface and, after traversing the surface potential in the reversed
direction, back to the plasma. Hence, for \SiOTwo, as well as any other
dielectric with mass mismatch $\overline{m}_e<1$, $S(E,\xi)={\cal T}(E,\xi)$ 
for $\xi < \sqrt{1-\overline{m}_e}$ and $S(E,\xi)<{\cal T}(E,\xi)$ for 
$\xi > \sqrt{1-\overline{m}_e}$. The sticking coefficient is thus only for 
some $E$ and $\xi$ equal to the transmission probability. This can be 
more clearly seen in the right panel of Fig.~\ref{AngleResolved}, where $S(E,\xi)$ 
(solid lines) and ${\cal T}(E,\xi)$ (dashed lines) are plotted as a function of $E$ for some 
representative $\xi$. To indicate the efficiency of our approach we mention that 
we obtained the about $3000$ data points for $S(E,\xi)$ in Fig.~\ref{AngleResolved}, 
corresponding each to a sum of trajectories with one backward and (depending on energy) up 
to $80$ forward scattering events, with the former interlaced between the latter in all 
possible ways, in only one hour computing time on a notebook. 

\begin{figure}[t]
\begin{minipage}{0.49\linewidth}
\includegraphics[width=0.85\linewidth]{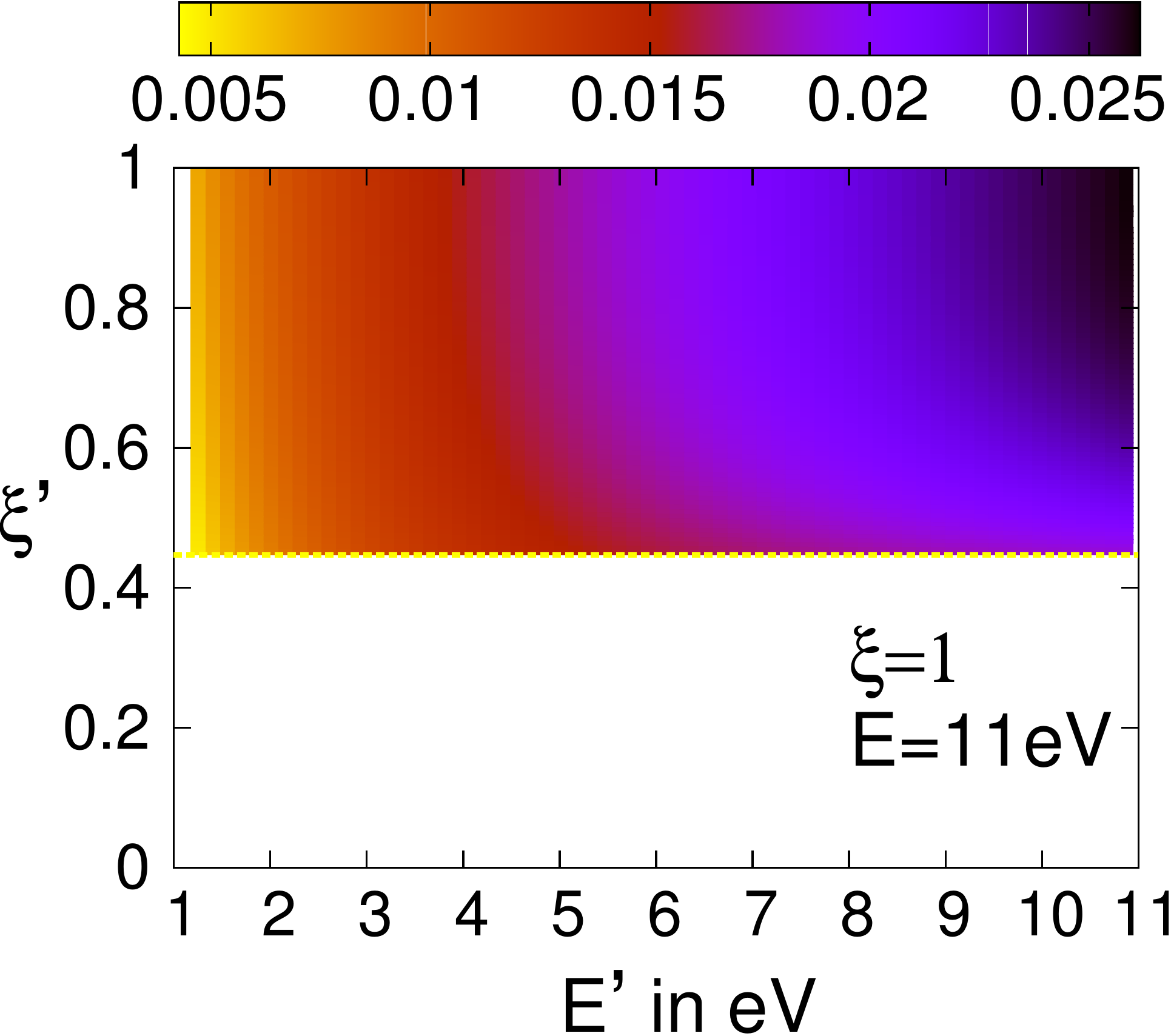}
\end{minipage}\begin{minipage}{0.5\linewidth}
\includegraphics[width=\linewidth]{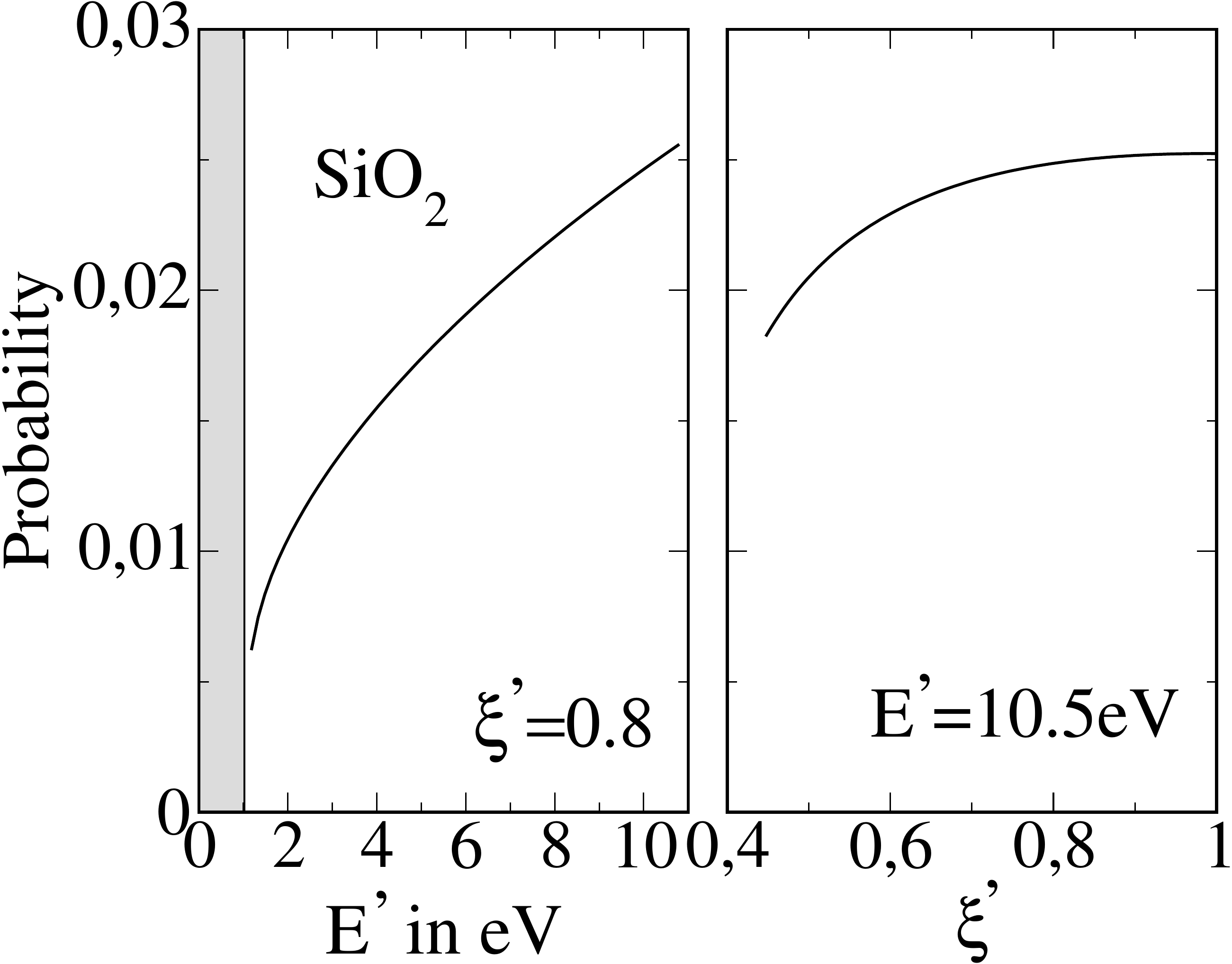}
\end{minipage}
\caption{(color online) On the left is plotted the probability $R_m^1(E; \eta(\xi)|\eta^\prime)$ 
for an electron hitting a clean \SiOTwo\ surface with $E=11\,{\rm eV}$ and $\xi=1$ to backscatter 
diffusely into a state with direction cosine $\xi^\prime=\xi(\eta^\prime)$ and energy $E^\prime$.
Below the dotted yellow line $R_m^1(E;\eta(\xi)|\eta^\prime)=0$ since diffuse backscattering 
cannot lead to post-collision direction cosines $\xi^\prime<\sqrt{1-\bar{m}_e}$. On the right 
are shown a horizontal and a vertical cut through the data depicted on the left. Diffuse 
backscattering peaks around the entrance energy $E=11\,{\rm eV}$ and the entrance direction 
cosine $\xi=1$.}
\label{BackScatt}
\end{figure}

The angle- and energy-resolved probability for diffuse backscattering $R_m^1(E;(\xi)|\eta^\prime)$
introduced in \eqref{Rm1} is depicted in Fig.~\ref{BackScatt} for $E=11\,{\rm eV}$ and
$\xi=1$. It is largest for $E^\prime=E$ and $\xi^\prime=\xi$. Post-collision 
direction cosines $\xi^\prime < \sqrt{1-\bar{m}_e}$ are excluded because mass mismatch 
and conservation of lateral momentum and total energy make them to correspond to 
internal states with perpendicular energy less than $\chi$. Scattering channels 
in these directions are thus closed. The maximum of the diffuse 
backscattering probability is always at the initial energy $E$ and 
direction cosine $\xi$. Had we chosen other values for $E$ and $\xi$ the plot would  
look similar only with a shifted maximum. 

Total reflection forces the sticking probability for an electron to vanish if it hits
the surface with energy $E$ and direction cosine $\xi<\xi_c$. It is caused by the mass 
mismatch and the conservation of lateral momentum and total energy. The former holds only 
for a homogeneous interface. In reality imperfections destroy the homogeneity. Lateral momentum 
will thus not be conserved and total reflection suppressed. To account for this possibility we 
now include elastic interface scattering along the lines Smith and coworkers used 
in their theoretical treatment of 
ballistic electron-emission spectroscopy~\cite{SLN98}.

Central to the approach is the probability for an electron hitting the wall from
the plasma with a kinetic energy $E-\chi>0$ to make a transition from $(E,\xi)$ to 
$(E^\prime,\xi^\prime)$ due to elastic scattering by any of the interfacial scattering 
centers, 
\begin{align}
P(E|E^\prime;\xi)=\frac{C/\xi}{1+C/\xi}\frac{\delta(E-E^\prime)}{\sqrt{E-\chi}}\theta(E-\chi)~,
\end{align}
where $C$ is a parameter proportional to the density of the scatterers and the 
square of the modulus of the scattering potential which we assume to be independent
of the initial and final scattering states (hard core scattering potential). Lacking a 
detailed knowledge of the structural properties of the interface we use $C$ as a fit 
parameter. The function $P(E|E^\prime;\xi)$ can be derived from the probability to 
make a transition due to scattering by a single center--given by the ratio of the golden 
rule scattering rate per time and the rate with which electrons hit the interface--taking  
interference corrections due to other centers into account~\cite{SLN98}. 

Any physical quantity $f(E,\xi)$ affected by interfacial disorder turns then into 
\begin{align}
\bar{f}&(E,\xi) = f(E,\xi)
\bigg[1-\int_0^1 d\xi^\prime\int_\chi^\infty \!\!\!\! dE^\prime \sqrt{E^\prime-\chi} P(E|E^\prime;\xi) \bigg] \nonumber\\
&+ \int_0^1 d\xi^\prime \int_\chi^\infty \!\!\!\! dE^\prime \sqrt{E^\prime-\chi}P(E|E^\prime;\xi) 
f(E^\prime,\xi^\prime)~,
\end{align}
where the first and second term on the rhs stand, respectively, for trajectories without 
and with interfacial scattering. Using this rule together with
\begin{align}
\int_0^1 d\xi^\prime \int_\chi^\infty \!\!\!\!dE^\prime \sqrt{E^\prime-\chi}\! P(E|E^\prime;\xi) = \frac{C/\xi}{1+C/\xi}
\end{align}	
yields for the sticking probability of a disordered dielectric wall~\cite{BF15}
\begin{align}
\bar{S}(E,\xi) &= \frac{{\cal T}(E,\xi)}{1+C/\xi}[1-\overline{\cal E}(E,\xi)] \nonumber\\
                    &+ \frac{C/\xi}{1+C/\xi} \int^1_{\xi_c} d\xi^\prime
                    T(E,\xi^\prime)[1-\bar{\cal E}(E,\xi^\prime)]~,
\label{Dirty}
\end{align}
where $\bar{\cal E}(E,\xi)$ is given by \eqref{EscapeProb} with 
${\cal T}(E,\xi)$ replaced by 
\begin{align}
\overline{\cal T}(E,\xi)=\frac{{\cal T}(E,\xi)}{1+C/\xi}  
                        + \frac{C/\xi}{1+C/\xi} \int^1_{\xi_c} d\xi^\prime {\cal T}(E,\xi^\prime)
\end{align}
and $\xi_c$ defined by~\eqref{xic}. Notice, in the limit $C\rightarrow 0$ we recover 
from \eqref{Dirty} the sticking probability $S(E,\xi)$ for a clean wall given by \eqref{StickCoeff} 
while for $C\rightarrow\infty$ we obtain the sticking probability for the totally disordered, dirty 
wall.

\begin{figure}[t]
\includegraphics[width=\linewidth]{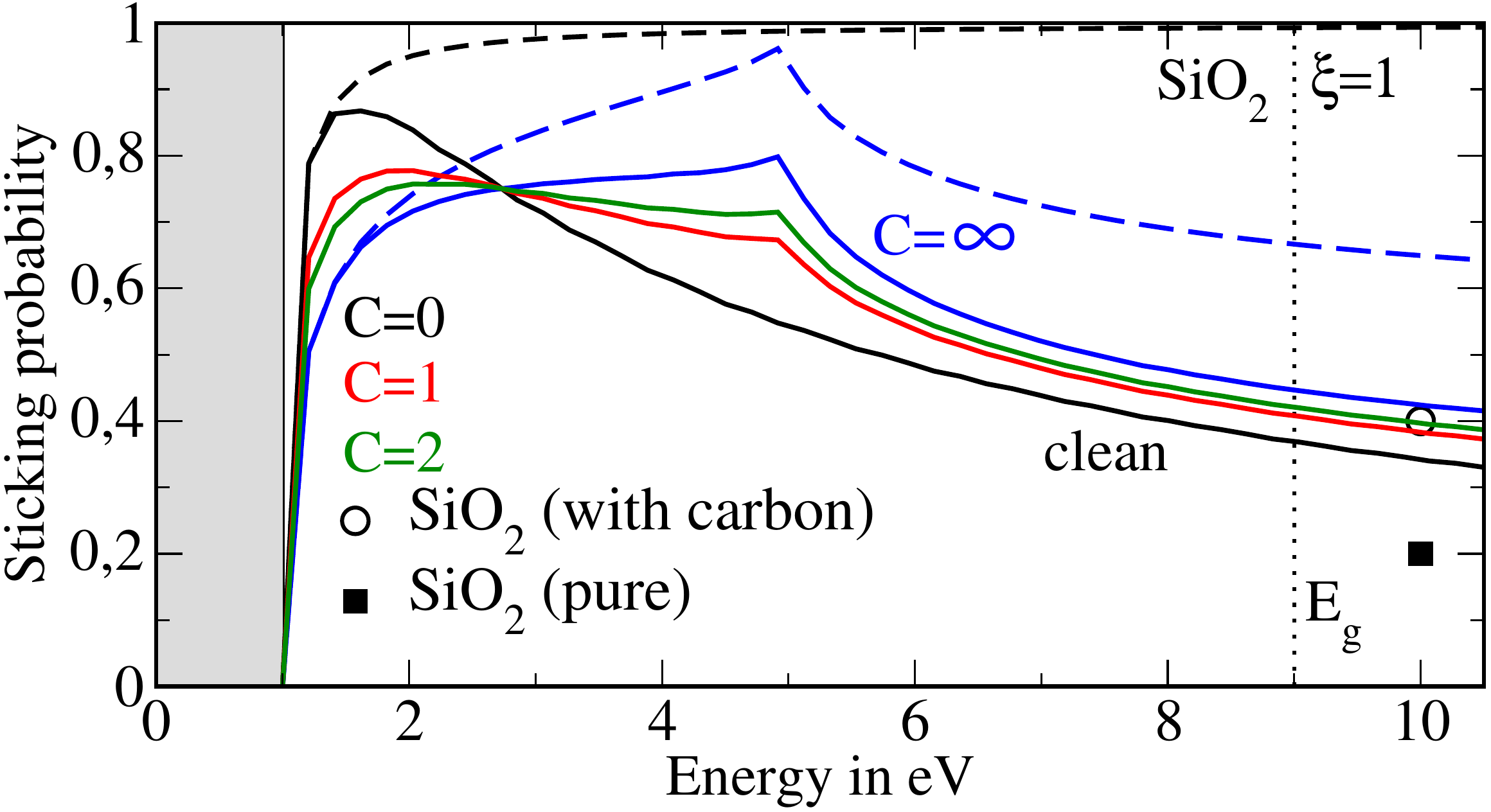}
\caption{(color online) Sticking probability $\overline{S}(E;1)$ for an electron
hitting a \SiOTwo\ surface perpendicularly as obtained from Eq.~\eqref{Dirty} (solid 
lines). We show data for $C=0$ (black), $1$ (red), $2$ (green), and $\infty$ (blue) 
with $\overline{\cal T}(E;1)$ also included for $C=0$ and $C=\infty$ (dashed lines).
Symbols are data from electron-beam scattering experiments~\cite{Dionne75}. 
}
\label{ExpData}
\end{figure}

The sticking probability $\bar{S}$ of a disordered \SiOTwo\ surface 
is shown in Fig.~\ref{ExpData} for $\xi=1$ (normal incident) and 
$C=0, 1, 2,$ and $\infty$. To indicate that our approach captures 
essential aspects of electron absorption by a surface we also plot 
data for two types of \SiOTwo\ surfaces obtained from electron-beam
scattering experiments~\cite{Dionne75}. Although the experimental data 
are in an energy range where electron-hole pair generation  
already starts to play a role they are nevertheless sufficiently close 
to the theoretical results to support our modeling approach. They 
also show that the perfect absorber value, $S(E,\xi)\approx 1$, is not 
applicable to \SiOTwo. For \MgO\ experimental data are available for lower 
energies showing a much better agreement with the theoretical data~\cite{BF15}. 

Dashed lines show, for comparison, $\overline{\cal T}(E,1)$, which is the 
sticking probability in the absence of backscattering. For $C=0$, 
$\bar{S}(E,1)$ deviates strongly from $\overline{\cal T}(E,1)$ (black lines), whereas 
for $C=\infty$ the two quantities approach each other (blue lines). The 
reason is the angle-averaging at the dirty surface (see Eq.~\eqref{Dirty}) 
which lessens, for a fixed $\xi$, the impact of inelastic backscattering compared
to the knock-out of propagation directions by total reflection. The kink in $\bar{S}(E,1)$ 
at $E=E_0$ signals the knock-out. Comparing the results for \MgO\ 
($\overline{m}_e=0.4$)~\cite{BF15} with the results for \SiOTwo\ ($\overline{m}_e=0.8$) 
indicates moreover that the closer $\overline{m}_e$ to unity the more affected is 
$\bar{S}(E,1)$ by inelastic backscattering. The mass mismatch $\overline{m}_e$ turns thus out 
to control $\bar{S}(E,1)$. This is not surprising. Because it is the effective electron mass 
which subsumes at low energy the elastic scattering of the electron by the ion cores of 
the wall.

\begin{table}[t]
\begin{center}
  \begin{tabular}{c|c|c|c|c|c|c}
 material     & $m_e^*/m_e$ & $\chi$[eV] & $\omega$[eV] & interface & $\tilde{s}$ & $Q_p[-e]$ \\\hline
    \SiOTwo   & 0.8 & 1.0 & 0.15 & clean  & 0.65 & 3028 \\
    \SiOTwo   & 0.8 & 1.0 & 0.15 & dirty  & 0.67 & 3063 \\\hline
\AlTwoOThree  & 0.4 & 2.5 & 0.10 & clean  & 0.42 & 2645 \\
\AlTwoOThree  & 0.4 & 2.5 & 0.10  & dirty  & 0.43 & 2645 \\\hline
        \MgO  & 0.4 & 1.0 & 0.10 & clean  & 0.28 & 2262 \\
        \MgO  & 0.4 & 1.0 & 0.10 & dirty  & 0.31 & 2367 \\\hline
PAM           & --  & -- & -- & --     & 1    & 3446 \\
  \end{tabular}
  \caption{Material parameters for the dielectrics used in this work and the orbital-motion
           limited charge $Q_p$ acquired in an argon plasma with $k_BT_e=2\,\rm{eV}$ and 
           $k_BT_i=0.026\,\rm{eV}$ by dust particles made out of these dielectrics and 
           having radius $R=1\,\mu{\rm m}$. The parameter $\tilde{s}$ defined in Eq.~\eqref{Soml}
           contains the material dependence of the particle charge.  
           As for \SiOTwo~\cite{BBS02,SF02,VMC11} the material parameters for 
           \MgO~\cite{KYA08,OGP03} and \AlTwoOThree~\cite{SWK03,BRB08} 
           can be different for actual particles/surfaces 
           due to material science aspects not addressed in this work. 
  }
  \label{OMLcharge}
\end{center}
\end{table}

At low energies, the electron-wall interaction in a plasma is usually treated within 
the perfect absorber model~\cite{Alpert65} stating that the probability with which an 
electron is absorbed (backscattered) by the wall is close to unity (vanishes). We 
have seen however that both probabilities are in fact energy- and angle-dependent
and deviate from the perfect absorber values.
It is thus of interest to work out the consequences of our results for the modeling
of bounded plasmas. As a first plasma application we consider the orbital-motion 
limited (OML) charging~\cite{Allen92} of a dielectric particle in a plasma. Similar
results would be however obtained for other charging models as well~\cite{LGS03,KRZ05}.

The grain surface is characterized by $\bar{S}(E,\xi)$ leading to an electron capture 
cross section $\sigma_c^e(E,\xi)=\bar{S}(E,\xi)\pi R^2(1+eV/E)$ with  
$E$ and $\xi$ the energy and direction cosine of the incident electron
and $V<0$ the grain's floating potential. The capture cross section yields the 
flux balance,
\begin{align}
\tilde{s}j_e^{\rm OML}(V)=j_i^{\rm OML}(V)~,
\label{FluxBalence}
\end{align}
where $j_e^{\rm OML}(V)$ and $j_i^{\rm OML}(V)$ are the OML  
fluxes~\cite{Allen92} and
\begin{align}
\tilde{s}=\int_\chi^\infty \!\! \frac{dE}{kT_e}
\exp\bigg[-\frac{E-\chi}{kT_e}\bigg]
\frac{E-\chi}{kT_e}\langle \bar{S}(E,\xi)\rangle_\xi
\label{Soml}
\end{align}
with
\begin{align}
\langle\bar{S}(E,\xi)\rangle_\xi=\int_0^1 d\xi\, \bar{S}(E,\xi)
\end{align}
the angle-averaged sticking probability. The (hard) perfect absorber assumption $\bar{S}(E,\xi)=1$ 
gives $\tilde{s}=1$.

\begin{figure}[t]
\includegraphics[width=\linewidth]{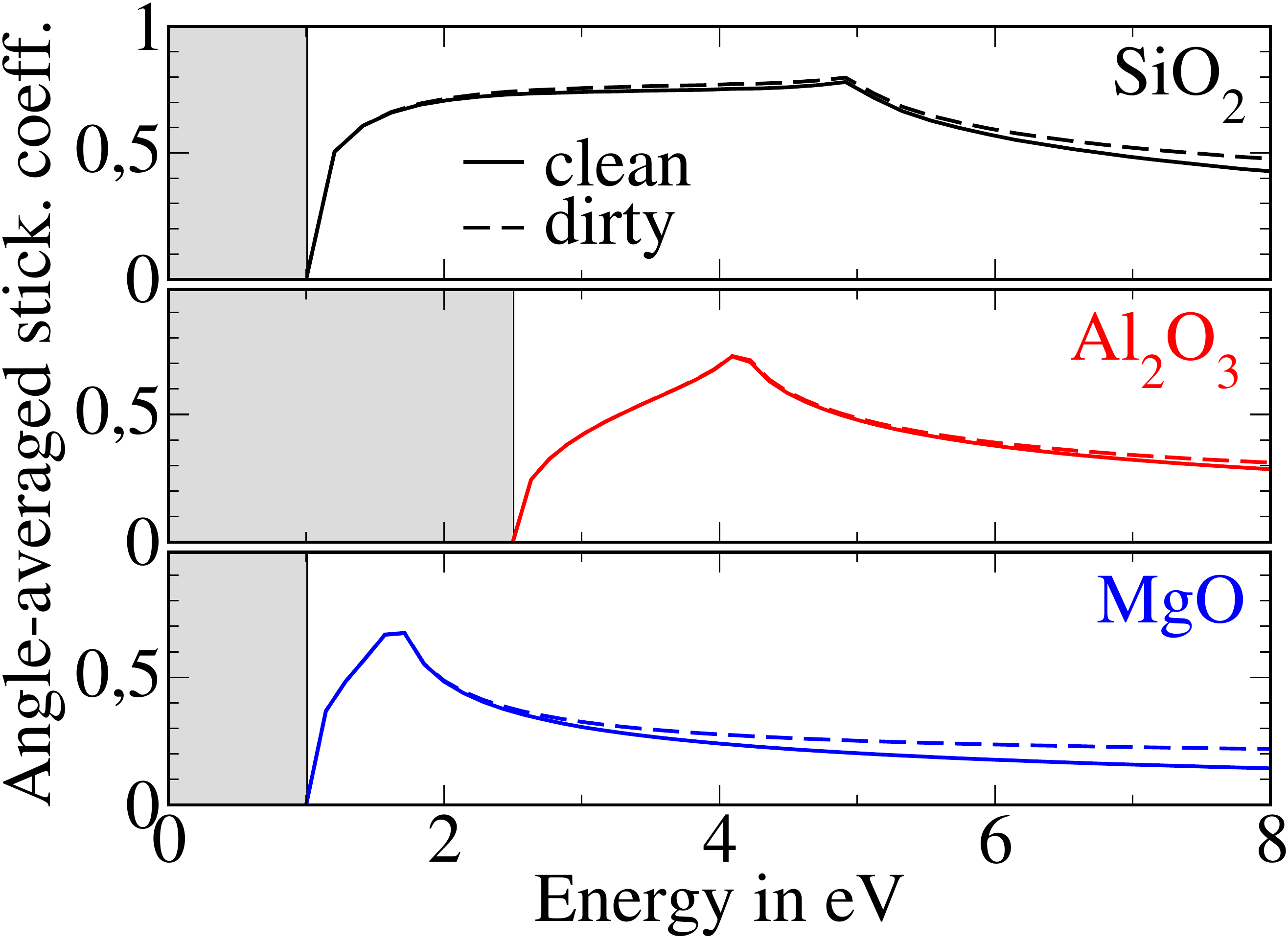}
\caption{Angle-averaged sticking coefficient $\langle \bar{S}(E,\xi)\rangle_\xi$ for a \SiOTwo, an 
\AlTwoOThree, and a \MgO\, surface. Solid and dashed lines indicate, respectively, 
$\langle \bar{S}(E,\xi)\rangle_\xi$ for a clean ($C=0$) and a dirty interface ($C=\infty$). The kink 
in the data signals the knock-out of propagation directions by total reflection. }
\label{AngleAve}
\end{figure}

In the limit $R\ll \lambda_D$ the grain charge in units of $-e$ is given by
$Q_p=-4\pi R\varepsilon_0kT_eV/e kT_e$
with $V$ the root of~\eqref{FluxBalence}. Table~\ref{OMLcharge} shows results for 
\SiOTwo, \AlTwoOThree, and \MgO. Besides the expected deviation of the grain charge 
from the perfect absorber (PAM) charge we also find a material dependence 
which could be tested by high precision charge measurements~\cite{CJG11}.
Figure~\ref{AngleAve} finally shows for \SiOTwo, \AlTwoOThree, and \MgO\, 
the angle-averaged sticking probability $\langle \bar{S}(E,\xi)\rangle_\xi$ 
entering~\eqref{Soml}. Since we calculate $\langle \bar{S}(E,\xi)\rangle_\xi$ 
only up to $E\simeq E_g$, for $E>E_g$ we should have included electron-hole pair 
generation leading to a more complicated recursion scheme containing energy 
integrals, we approximate in~\eqref{Soml} $\langle \bar{S}(E,\xi)\rangle_\xi$ 
for $E>E_g$ by its upper bound $\langle \overline{\cal T}(E,\xi)\rangle_\xi$. The 
values for $\tilde{s}$ given in Table~\ref{OMLcharge} are thus only upper bounds.

\section{Conclusion}
\label{Conclusions}

We described a method to calculate from a microscopic model the sticking and 
backscattering probabilities for a low-energy electron hitting the wall of a plasma.
Taking advantage of the large penetration depth at low energies the method factorizes 
sticking and backscattering probabilities into probabilities for quantum-mechanical 
transmission and internal backscattering. For the description of the latter 
we employed an invariant embedding principle. It allows us to extract from the
great number of electron trajectories the most important ones which are 
the trajectories bringing the electron back to the plasma after at least one 
backscattering but as many forward scattering events as are energetically possible.
The approach is applicable to metallic as well as dielectric surfaces. 
For dielectric surfaces at room temperature, where emission of optical phonons is 
the dominant scattering process, it is sufficient to include only one 
backscattering event interlaced in all possible ways between 
the energetically allowed sequence of forward scattering events.

In this work we focused on \SiOTwo\ and found good agreement with experimental 
data from electron-beam scattering. Contrary to the perfect absorber assumption, 
$\bar{S}(E,\xi)\approx 1$, we find energy- and angle-dependent sticking probabilities 
which can deviate significantly from unity because of internal backscattering 
due to emission of optical phonons and total reflection due to electron mass mismatch
and conservation of total energy and lateral momentum. Angle-averaged sticking 
probabilities $\langle \bar{S}(E,\xi)\rangle_\xi$ for \SiOTwo, \AlTwoOThree, and \MgO\ 
are also less than unity. Incorporating the sticking probabilities into orbital-motion 
limited charging fluxes reduces the grain charge by about 10 percent compared to the 
perfect absorber value and makes the charge material-dependent.

The method is particularly strong for electron energies below $100\,{\rm eV}$. 
Various scattering mechanism can be included as well as imperfections of the interface.
It could thus be used to systematically investigate the interaction of electrons with
the walls of low-temperature plasmas. This is indeed required. Electron sticking 
and backscattering are not universal in the energy range of interest for plasma
applications. They have to be studied for each wall material separately.

We acknowledge support by the Deutsche Forschungsgemeinschaft through 
SFB/TRR 24.


\begin{thebibliography}{52}
\expandafter\ifx\csname natexlab\endcsname\relax\def\natexlab#1{#1}\fi
\expandafter\ifx\csname bibnamefont\endcsname\relax
  \def\bibnamefont#1{#1}\fi
\expandafter\ifx\csname bibfnamefont\endcsname\relax
  \def\bibfnamefont#1{#1}\fi
\expandafter\ifx\csname citenamefont\endcsname\relax
  \def\citenamefont#1{#1}\fi
\expandafter\ifx\csname url\endcsname\relax
  \def\url#1{\texttt{#1}}\fi
\expandafter\ifx\csname urlprefix\endcsname\relax\def\urlprefix{URL }\fi
\providecommand{\bibinfo}[2]{#2}
\providecommand{\eprint}[2][]{\url{#2}}

\bibitem[{\citenamefont{Mott-Smith and Langmuir}(1926)}]{MSL26}
\bibinfo{author}{\bibfnamefont{H.~M.} \bibnamefont{Mott-Smith}}
  \bibnamefont{and} \bibinfo{author}{\bibfnamefont{I.}~\bibnamefont{Langmuir}},
  \bibinfo{journal}{Phys. Rev.} \textbf{\bibinfo{volume}{28}},
  \bibinfo{pages}{727} (\bibinfo{year}{1926}).

\bibitem[{\citenamefont{Franklin}(1976)}]{Franklin76}
\bibinfo{author}{\bibfnamefont{R.~N.} \bibnamefont{Franklin}},
  \emph{\bibinfo{title}{Plasma phenomena in gas discharges}}
  (\bibinfo{publisher}{Clarendon Press}, \bibinfo{address}{Oxford},
  \bibinfo{year}{1976}).

\bibitem[{\citenamefont{Tolias}(2014)}]{Tolias14a}
\bibinfo{author}{\bibfnamefont{P.}~\bibnamefont{Tolias}},
  \bibinfo{journal}{Plasma Phys. Control. Fusion}
  \textbf{\bibinfo{volume}{56}}, \bibinfo{pages}{123002}
  (\bibinfo{year}{2014}).

\bibitem[{\citenamefont{Bogaczyk et~al.}(2012)\citenamefont{Bogaczyk, Wild,
  Stollenwerk, and Wagner}}]{BWS12}
\bibinfo{author}{\bibfnamefont{M.}~\bibnamefont{Bogaczyk}},
  \bibinfo{author}{\bibfnamefont{R.}~\bibnamefont{Wild}},
  \bibinfo{author}{\bibfnamefont{L.}~\bibnamefont{Stollenwerk}},
  \bibnamefont{and} \bibinfo{author}{\bibfnamefont{H.-E.}
  \bibnamefont{Wagner}}, \bibinfo{journal}{J. Phys. D: Appl. Phys.}
  \textbf{\bibinfo{volume}{45}}, \bibinfo{pages}{465202}
  (\bibinfo{year}{2012}).

\bibitem[{\citenamefont{Tschiersch et~al.}(2014)\citenamefont{Tschiersch,
  Bogaczyk, and Wagner}}]{TBW14}
\bibinfo{author}{\bibfnamefont{R.}~\bibnamefont{Tschiersch}},
  \bibinfo{author}{\bibfnamefont{M.}~\bibnamefont{Bogaczyk}}, \bibnamefont{and}
  \bibinfo{author}{\bibfnamefont{H.-E.} \bibnamefont{Wagner}},
  \bibinfo{journal}{J. Phys. D: Appl. Phys.} \textbf{\bibinfo{volume}{47}},
  \bibinfo{pages}{365204} (\bibinfo{year}{2014}).

\bibitem[{\citenamefont{Peeters and van~de Sanden}(2015)}]{PS15}
\bibinfo{author}{\bibfnamefont{F.~J.~J.} \bibnamefont{Peeters}}
  \bibnamefont{and} \bibinfo{author}{\bibfnamefont{M.~C.~M.}
  \bibnamefont{van~de Sanden}}, \bibinfo{journal}{Plasma Sources Sci. Technol.}
  \textbf{\bibinfo{volume}{24}}, \bibinfo{pages}{015016}
  (\bibinfo{year}{2015}).

\bibitem[{\citenamefont{Richterov\'a et~al.}(2010)\citenamefont{Richterov\'a,
  Ber\'anek, Pavl\r{u}, N\v{e}me\v{c}ek, and \v{S}\'afrankov\'a}}]{RBP10}
\bibinfo{author}{\bibfnamefont{I.}~\bibnamefont{Richterov\'a}},
  \bibinfo{author}{\bibfnamefont{M.}~\bibnamefont{Ber\'anek}},
  \bibinfo{author}{\bibfnamefont{J.}~\bibnamefont{Pavl\r{u}}},
  \bibinfo{author}{\bibfnamefont{Z.}~\bibnamefont{N\v{e}me\v{c}ek}},
  \bibnamefont{and}
  \bibinfo{author}{\bibfnamefont{J.}~\bibnamefont{\v{S}\'afrankov\'a}},
  \bibinfo{journal}{Phys. Rev. B} \textbf{\bibinfo{volume}{81}},
  \bibinfo{pages}{075406} (\bibinfo{year}{2010}).

\bibitem[{\citenamefont{Ishihara}(2007)}]{Ishihara07}
\bibinfo{author}{\bibfnamefont{O.}~\bibnamefont{Ishihara}},
  \bibinfo{journal}{J. Phys. D: Appl. Phys.} \textbf{\bibinfo{volume}{40}},
  \bibinfo{pages}{R121} (\bibinfo{year}{2007}).

\bibitem[{\citenamefont{Fortov et~al.}(2005)\citenamefont{Fortov, Ivlev,
  Khrapak, Khrapak, and Morfill}}]{FIK05}
\bibinfo{author}{\bibfnamefont{V.~E.} \bibnamefont{Fortov}},
  \bibinfo{author}{\bibfnamefont{A.~V.} \bibnamefont{Ivlev}},
  \bibinfo{author}{\bibfnamefont{S.~A.} \bibnamefont{Khrapak}},
  \bibinfo{author}{\bibfnamefont{A.~G.} \bibnamefont{Khrapak}},
  \bibnamefont{and} \bibinfo{author}{\bibfnamefont{G.~E.}
  \bibnamefont{Morfill}}, \bibinfo{journal}{Phys. Rep.}
  \textbf{\bibinfo{volume}{421}}, \bibinfo{pages}{1} (\bibinfo{year}{2005}).

\bibitem[{\citenamefont{Dunaevsky et~al.}(2003)\citenamefont{Dunaevsky,
  Raitses, and Fisch}}]{DRF03}
\bibinfo{author}{\bibfnamefont{A.}~\bibnamefont{Dunaevsky}},
  \bibinfo{author}{\bibfnamefont{Y.}~\bibnamefont{Raitses}}, \bibnamefont{and}
  \bibinfo{author}{\bibfnamefont{N.~J.} \bibnamefont{Fisch}},
  \bibinfo{journal}{Phys. Plasmas} \textbf{\bibinfo{volume}{10}},
  \bibinfo{pages}{2574} (\bibinfo{year}{2003}).

\bibitem[{\citenamefont{Barral et~al.}(2003)\citenamefont{Barral, Makowski,
  Peradzy\'nski, Gascon, and Dudeck}}]{BMP03}
\bibinfo{author}{\bibfnamefont{S.}~\bibnamefont{Barral}},
  \bibinfo{author}{\bibfnamefont{K.}~\bibnamefont{Makowski}},
  \bibinfo{author}{\bibfnamefont{Z.}~\bibnamefont{Peradzy\'nski}},
  \bibinfo{author}{\bibfnamefont{N.}~\bibnamefont{Gascon}}, \bibnamefont{and}
  \bibinfo{author}{\bibfnamefont{M.}~\bibnamefont{Dudeck}},
  \bibinfo{journal}{Phys. Plasmas} \textbf{\bibinfo{volume}{10}},
  \bibinfo{pages}{4137} (\bibinfo{year}{2003}).

\bibitem[{\citenamefont{Lapke et~al.}(2007)\citenamefont{Lapke, Mussenbrock,
  Brinkmann, Scharwitz, Boeke, and Winter}}]{LMB07}
\bibinfo{author}{\bibfnamefont{M.}~\bibnamefont{Lapke}},
  \bibinfo{author}{\bibfnamefont{T.}~\bibnamefont{Mussenbrock}},
  \bibinfo{author}{\bibfnamefont{R.~P.} \bibnamefont{Brinkmann}},
  \bibinfo{author}{\bibfnamefont{C.}~\bibnamefont{Scharwitz}},
  \bibinfo{author}{\bibfnamefont{M.}~\bibnamefont{Boeke}}, \bibnamefont{and}
  \bibinfo{author}{\bibfnamefont{J.}~\bibnamefont{Winter}},
  \bibinfo{journal}{Appl. Phys. Lett.} \textbf{\bibinfo{volume}{90}},
  \bibinfo{pages}{121502} (\bibinfo{year}{2007}).

\bibitem[{\citenamefont{Cazaux}(2012)}]{Cazaux12}
\bibinfo{author}{\bibfnamefont{J.}~\bibnamefont{Cazaux}}, \bibinfo{journal}{J.
  Appl. Phys.} \textbf{\bibinfo{volume}{111}}, \bibinfo{pages}{064903}
  (\bibinfo{year}{2012}).

\bibitem[{\citenamefont{Salvat-Pujol and Werner}(2013)}]{SW13}
\bibinfo{author}{\bibfnamefont{F.}~\bibnamefont{Salvat-Pujol}}
  \bibnamefont{and} \bibinfo{author}{\bibfnamefont{W.~S.~M.}
  \bibnamefont{Werner}}, \bibinfo{journal}{Surf. Interface Anal.}
  \textbf{\bibinfo{volume}{45}}, \bibinfo{pages}{873} (\bibinfo{year}{2013}).

\bibitem[{\citenamefont{Werner}(2001)}]{Werner01}
\bibinfo{author}{\bibfnamefont{W.~S.~M.} \bibnamefont{Werner}},
  \bibinfo{journal}{Surf. Interface Anal.} \textbf{\bibinfo{volume}{31}},
  \bibinfo{pages}{141} (\bibinfo{year}{2001}).

\bibitem[{\citenamefont{Klassen et~al.}(2014)\citenamefont{Klassen, Bauereiss,
  and Koerner}}]{KBK14}
\bibinfo{author}{\bibfnamefont{A.}~\bibnamefont{Klassen}},
  \bibinfo{author}{\bibfnamefont{A.}~\bibnamefont{Bauereiss}},
  \bibnamefont{and} \bibinfo{author}{\bibfnamefont{C.}~\bibnamefont{Koerner}},
  \bibinfo{journal}{J. Phys. D: Appl. Phys.} \textbf{\bibinfo{volume}{47}},
  \bibinfo{pages}{065307} (\bibinfo{year}{2014}).

\bibitem[{\citenamefont{Glazov and Tougaard}(2003)}]{GT03}
\bibinfo{author}{\bibfnamefont{L.~G.} \bibnamefont{Glazov}} \bibnamefont{and}
  \bibinfo{author}{\bibfnamefont{S.}~\bibnamefont{Tougaard}},
  \bibinfo{journal}{Phys. Rev. B} \textbf{\bibinfo{volume}{68}},
  \bibinfo{pages}{155409} (\bibinfo{year}{2003}).

\bibitem[{\citenamefont{Schreiber and Fitting}(2002)}]{SF02}
\bibinfo{author}{\bibfnamefont{E.}~\bibnamefont{Schreiber}} \bibnamefont{and}
  \bibinfo{author}{\bibfnamefont{H.-J.} \bibnamefont{Fitting}},
  \bibinfo{journal}{J. Electron Spectros. Relat. Phenom.}
  \textbf{\bibinfo{volume}{124}}, \bibinfo{pages}{25} (\bibinfo{year}{2002}).

\bibitem[{\citenamefont{Dubus et~al.}(2000)\citenamefont{Dubus, Jablonski, and
  Tougaard}}]{DJT00}
\bibinfo{author}{\bibfnamefont{A.}~\bibnamefont{Dubus}},
  \bibinfo{author}{\bibfnamefont{A.}~\bibnamefont{Jablonski}},
  \bibnamefont{and} \bibinfo{author}{\bibfnamefont{S.}~\bibnamefont{Tougaard}},
  \bibinfo{journal}{Progr. Surface Science} \textbf{\bibinfo{volume}{63}},
  \bibinfo{pages}{135} (\bibinfo{year}{2000}).

\bibitem[{\citenamefont{Vicanek}(1999)}]{Vicanek99}
\bibinfo{author}{\bibfnamefont{M.}~\bibnamefont{Vicanek}},
  \bibinfo{journal}{Surface Science} \textbf{\bibinfo{volume}{440}},
  \bibinfo{pages}{1} (\bibinfo{year}{1999}).

\bibitem[{\citenamefont{Dudarev et~al.}(1995)\citenamefont{Dudarev, Rez, and
  Whelan}}]{DRW95}
\bibinfo{author}{\bibfnamefont{S.~L.} \bibnamefont{Dudarev}},
  \bibinfo{author}{\bibfnamefont{P.}~\bibnamefont{Rez}}, \bibnamefont{and}
  \bibinfo{author}{\bibfnamefont{M.~J.} \bibnamefont{Whelan}},
  \bibinfo{journal}{Phys. Rev. B} \textbf{\bibinfo{volume}{51}},
  \bibinfo{pages}{3397} (\bibinfo{year}{1995}).

\bibitem[{\citenamefont{Tofterup}(1985)}]{Tofterup85}
\bibinfo{author}{\bibfnamefont{A.~L.} \bibnamefont{Tofterup}},
  \bibinfo{journal}{Phys. Rev. B} \textbf{\bibinfo{volume}{32}},
  \bibinfo{pages}{2808} (\bibinfo{year}{1985}).

\bibitem[{\citenamefont{Kanaya and Okayama}(1972)}]{KO72}
\bibinfo{author}{\bibfnamefont{K.}~\bibnamefont{Kanaya}} \bibnamefont{and}
  \bibinfo{author}{\bibfnamefont{S.}~\bibnamefont{Okayama}},
  \bibinfo{journal}{J. Phys. D: Appl. Phys.} \textbf{\bibinfo{volume}{5}},
  \bibinfo{pages}{43} (\bibinfo{year}{1972}).

\bibitem[{\citenamefont{Dashen}(1964)}]{Dashen64}
\bibinfo{author}{\bibfnamefont{R.}~\bibnamefont{Dashen}},
  \bibinfo{journal}{Phys. Rev.} \textbf{\bibinfo{volume}{134}},
  \bibinfo{pages}{A1025} (\bibinfo{year}{1964}).

\bibitem[{\citenamefont{Shihab et~al.}(2013)\citenamefont{Shihab, Elgendy,
  Korolov, Derzsi, Schulze, Eremin, Mussenbrock, Donk\'o, and
  Brinkmann}}]{SEK13}
\bibinfo{author}{\bibfnamefont{M.}~\bibnamefont{Shihab}},
  \bibinfo{author}{\bibfnamefont{A.~T.} \bibnamefont{Elgendy}},
  \bibinfo{author}{\bibfnamefont{I.}~\bibnamefont{Korolov}},
  \bibinfo{author}{\bibfnamefont{A.}~\bibnamefont{Derzsi}},
  \bibinfo{author}{\bibfnamefont{J.}~\bibnamefont{Schulze}},
  \bibinfo{author}{\bibfnamefont{D.}~\bibnamefont{Eremin}},
  \bibinfo{author}{\bibfnamefont{T.}~\bibnamefont{Mussenbrock}},
  \bibinfo{author}{\bibfnamefont{Z.}~\bibnamefont{Donk\'o}}, \bibnamefont{and}
  \bibinfo{author}{\bibfnamefont{R.~P.} \bibnamefont{Brinkmann}},
  \bibinfo{journal}{Plasma Sources Sci. Technol.}
  \textbf{\bibinfo{volume}{22}}, \bibinfo{pages}{055013}
  (\bibinfo{year}{2013}).

\bibitem[{\citenamefont{Stollenwerk et~al.}(2007)\citenamefont{Stollenwerk,
  Amiranashvili, Boeuf, and Purwins}}]{SAB07}
\bibinfo{author}{\bibfnamefont{L.}~\bibnamefont{Stollenwerk}},
  \bibinfo{author}{\bibfnamefont{S.}~\bibnamefont{Amiranashvili}},
  \bibinfo{author}{\bibfnamefont{J.-P.} \bibnamefont{Boeuf}}, \bibnamefont{and}
  \bibinfo{author}{\bibfnamefont{H.-G.} \bibnamefont{Purwins}},
  \bibinfo{journal}{Eur. Phys. J. D} \textbf{\bibinfo{volume}{44}},
  \bibinfo{pages}{133} (\bibinfo{year}{2007}).

\bibitem[{\citenamefont{Kersten et~al.}(2004)\citenamefont{Kersten, Deutsch,
  and Kroesen}}]{KDK04}
\bibinfo{author}{\bibfnamefont{H.}~\bibnamefont{Kersten}},
  \bibinfo{author}{\bibfnamefont{H.}~\bibnamefont{Deutsch}}, \bibnamefont{and}
  \bibinfo{author}{\bibfnamefont{G.~M.~W.} \bibnamefont{Kroesen}},
  \bibinfo{journal}{Int. J. Mass Spectr.} \textbf{\bibinfo{volume}{233}},
  \bibinfo{pages}{51} (\bibinfo{year}{2004}).

\bibitem[{\citenamefont{Brandenburg et~al.}(2005)\citenamefont{Brandenburg,
  Maiorov, Golubovskii, Wagner, Behnke, and Behnke}}]{BMG05}
\bibinfo{author}{\bibfnamefont{R.}~\bibnamefont{Brandenburg}},
  \bibinfo{author}{\bibfnamefont{V.~A.} \bibnamefont{Maiorov}},
  \bibinfo{author}{\bibfnamefont{Y.~B.} \bibnamefont{Golubovskii}},
  \bibinfo{author}{\bibfnamefont{H.-E.} \bibnamefont{Wagner}},
  \bibinfo{author}{\bibfnamefont{J.}~\bibnamefont{Behnke}}, \bibnamefont{and}
  \bibinfo{author}{\bibfnamefont{J.~F.} \bibnamefont{Behnke}},
  \bibinfo{journal}{J. Phys. D: Appl. Phys.} \textbf{\bibinfo{volume}{38}},
  \bibinfo{pages}{2187} (\bibinfo{year}{2005}).

\bibitem[{\citenamefont{Uhrlandt et~al.}(2000)\citenamefont{Uhrlandt, Schmidt,
  Behnke, and Bindemann}}]{USB00}
\bibinfo{author}{\bibfnamefont{D.}~\bibnamefont{Uhrlandt}},
  \bibinfo{author}{\bibfnamefont{M.}~\bibnamefont{Schmidt}},
  \bibinfo{author}{\bibfnamefont{J.~F.} \bibnamefont{Behnke}},
  \bibnamefont{and}
  \bibinfo{author}{\bibfnamefont{T.}~\bibnamefont{Bindemann}},
  \bibinfo{journal}{J. Phys. D: Appl. Phys.} \textbf{\bibinfo{volume}{33}},
  \bibinfo{pages}{2475} (\bibinfo{year}{2000}).

\bibitem[{\citenamefont{Al'pert et~al.}(1965)\citenamefont{Al'pert, Gurevich,
  and Pitaevskii}}]{Alpert65}
\bibinfo{author}{\bibfnamefont{Y.~L.} \bibnamefont{Al'pert}},
  \bibinfo{author}{\bibfnamefont{A.~V.} \bibnamefont{Gurevich}},
  \bibnamefont{and} \bibinfo{author}{\bibfnamefont{L.~P.}
  \bibnamefont{Pitaevskii}}, \emph{\bibinfo{title}{Space physics with
  artificial satellites}} (\bibinfo{publisher}{Consultants Bureau},
  \bibinfo{address}{New York}, \bibinfo{year}{1965}).

\bibitem[{\citenamefont{Mendis}(2002)}]{Mendis02}
\bibinfo{author}{\bibfnamefont{D.~A.} \bibnamefont{Mendis}},
  \bibinfo{journal}{Plasma Sources Sci. Technol.}
  \textbf{\bibinfo{volume}{11}}, \bibinfo{pages}{A219} (\bibinfo{year}{2002}).

\bibitem[{\citenamefont{Bronold and Fehske}(2015)}]{BF15}
\bibinfo{author}{\bibfnamefont{F.~X.} \bibnamefont{Bronold}} \bibnamefont{and}
  \bibinfo{author}{\bibfnamefont{H.}~\bibnamefont{Fehske}},
  \bibinfo{journal}{Phys. Rev. Lett.} \textbf{\bibinfo{volume}{115}},
  \bibinfo{pages}{225001} (\bibinfo{year}{2015}).

\bibitem[{\citenamefont{Hickmott}(1965)}]{Hickmott65}
\bibinfo{author}{\bibfnamefont{T.~W.} \bibnamefont{Hickmott}},
  \bibinfo{journal}{J. Appl. Phys.} \textbf{\bibinfo{volume}{36}},
  \bibinfo{pages}{1885} (\bibinfo{year}{1965}).

\bibitem[{\citenamefont{Glazov and P\'azsit}(2007)}]{GP07}
\bibinfo{author}{\bibfnamefont{L.~G.} \bibnamefont{Glazov}} \bibnamefont{and}
  \bibinfo{author}{\bibfnamefont{I.}~\bibnamefont{P\'azsit}},
  \bibinfo{journal}{Nucl. Instr. and Meth. B} \textbf{\bibinfo{volume}{256}},
  \bibinfo{pages}{638} (\bibinfo{year}{2007}).

\bibitem[{\citenamefont{Cook and Fredericks}(1962)}]{CF62}
\bibinfo{author}{\bibfnamefont{C.~J.} \bibnamefont{Cook}} \bibnamefont{and}
  \bibinfo{author}{\bibfnamefont{W.~J.} \bibnamefont{Fredericks}},
  \bibinfo{journal}{J. Chem. Phys} \textbf{\bibinfo{volume}{36}},
  \bibinfo{pages}{608} (\bibinfo{year}{1962}).

\bibitem[{\citenamefont{Dionne}(1975)}]{Dionne75}
\bibinfo{author}{\bibfnamefont{G.~F.} \bibnamefont{Dionne}},
  \bibinfo{journal}{J. Appl. Phys.} \textbf{\bibinfo{volume}{46}},
  \bibinfo{pages}{3347} (\bibinfo{year}{1975}).

\bibitem[{\citenamefont{Heinisch et~al.}(2012)\citenamefont{Heinisch, Bronold,
  and Fehske}}]{HBF12}
\bibinfo{author}{\bibfnamefont{R.~L.} \bibnamefont{Heinisch}},
  \bibinfo{author}{\bibfnamefont{F.~X.} \bibnamefont{Bronold}},
  \bibnamefont{and} \bibinfo{author}{\bibfnamefont{H.}~\bibnamefont{Fehske}},
  \bibinfo{journal}{Phys. Rev. B} \textbf{\bibinfo{volume}{85}},
  \bibinfo{pages}{075323} (\bibinfo{year}{2012}).

\bibitem[{\citenamefont{Wu and Yang}(1979)}]{WY79}
\bibinfo{author}{\bibfnamefont{C.~M.} \bibnamefont{Wu}} \bibnamefont{and}
  \bibinfo{author}{\bibfnamefont{E.~S.} \bibnamefont{Yang}},
  \bibinfo{journal}{Solid-State Electronics} \textbf{\bibinfo{volume}{22}},
  \bibinfo{pages}{241} (\bibinfo{year}{1979}).

\bibitem[{\citenamefont{Gaylord and Brennan}(1989)}]{GB89}
\bibinfo{author}{\bibfnamefont{T.~K.} \bibnamefont{Gaylord}} \bibnamefont{and}
  \bibinfo{author}{\bibfnamefont{K.~F.} \bibnamefont{Brennan}},
  \bibinfo{journal}{J. Appl. Phys.} \textbf{\bibinfo{volume}{65}},
  \bibinfo{pages}{814} (\bibinfo{year}{1989}).

\bibitem[{\citenamefont{Ridley}(1998)}]{Ridley98}
\bibinfo{author}{\bibfnamefont{B.~K.} \bibnamefont{Ridley}},
  \bibinfo{journal}{J. Appl. Phys.} \textbf{\bibinfo{volume}{84}},
  \bibinfo{pages}{4020} (\bibinfo{year}{1998}).

\bibitem[{\citenamefont{Oswald et~al.}(1993)\citenamefont{Oswald, Kasper, and
  Gaukler}}]{OKG93}
\bibinfo{author}{\bibfnamefont{R.}~\bibnamefont{Oswald}},
  \bibinfo{author}{\bibfnamefont{E.}~\bibnamefont{Kasper}}, \bibnamefont{and}
  \bibinfo{author}{\bibfnamefont{K.~H.} \bibnamefont{Gaukler}},
  \bibinfo{journal}{J. Electron Spectros. Relat. Phenom.}
  \textbf{\bibinfo{volume}{61}}, \bibinfo{pages}{251} (\bibinfo{year}{1993}).

\bibitem[{\citenamefont{Ballarotto et~al.}(2002)\citenamefont{Ballarotto,
  Breban, Siegrist, Phaneuf, and Williams}}]{BBS02}
\bibinfo{author}{\bibfnamefont{V.~W.} \bibnamefont{Ballarotto}},
  \bibinfo{author}{\bibfnamefont{M.}~\bibnamefont{Breban}},
  \bibinfo{author}{\bibfnamefont{K.}~\bibnamefont{Siegrist}},
  \bibinfo{author}{\bibfnamefont{R.~J.} \bibnamefont{Phaneuf}},
  \bibnamefont{and} \bibinfo{author}{\bibfnamefont{E.~D.}
  \bibnamefont{Williams}}, \bibinfo{journal}{J. Vac. Sci. Technol. B}
  \textbf{\bibinfo{volume}{20}}, \bibinfo{pages}{2514} (\bibinfo{year}{2002}).

\bibitem[{\citenamefont{Vella et~al.}(2011)\citenamefont{Vella, Messina,
  Cannas, and Boscaino}}]{VMC11}
\bibinfo{author}{\bibfnamefont{E.}~\bibnamefont{Vella}},
  \bibinfo{author}{\bibfnamefont{F.}~\bibnamefont{Messina}},
  \bibinfo{author}{\bibfnamefont{M.}~\bibnamefont{Cannas}}, \bibnamefont{and}
  \bibinfo{author}{\bibfnamefont{R.}~\bibnamefont{Boscaino}},
  \bibinfo{journal}{Phys. Rev. B} \textbf{\bibinfo{volume}{83}},
  \bibinfo{pages}{174201} (\bibinfo{year}{2011}).

\bibitem[{\citenamefont{Smith et~al.}(1998)\citenamefont{Smith, Lee, and
  Narayanamurti}}]{SLN98}
\bibinfo{author}{\bibfnamefont{D.~L.} \bibnamefont{Smith}},
  \bibinfo{author}{\bibfnamefont{E.~Y.} \bibnamefont{Lee}}, \bibnamefont{and}
  \bibinfo{author}{\bibfnamefont{V.}~\bibnamefont{Narayanamurti}},
  \bibinfo{journal}{Phys. Rev. Lett.} \textbf{\bibinfo{volume}{80}},
  \bibinfo{pages}{2433} (\bibinfo{year}{1998}).

\bibitem[{\citenamefont{Kim et~al.}(2008)\citenamefont{Kim, Yoon, Ahn, Hong,
  and Yang}}]{KYA08}
\bibinfo{author}{\bibfnamefont{Y.-S.} \bibnamefont{Kim}},
  \bibinfo{author}{\bibfnamefont{S.-H.} \bibnamefont{Yoon}},
  \bibinfo{author}{\bibfnamefont{S.-G.} \bibnamefont{Ahn}},
  \bibinfo{author}{\bibfnamefont{C.-R.} \bibnamefont{Hong}}, \bibnamefont{and}
  \bibinfo{author}{\bibfnamefont{H.}~\bibnamefont{Yang}},
  \bibinfo{journal}{Electron. Mater. Lett.} \textbf{\bibinfo{volume}{4}},
  \bibinfo{pages}{113} (\bibinfo{year}{2008}).

\bibitem[{\citenamefont{Oganov et~al.}(2003)\citenamefont{Oganov, Gillan, and
  Price}}]{OGP03}
\bibinfo{author}{\bibfnamefont{A.~R.} \bibnamefont{Oganov}},
  \bibinfo{author}{\bibfnamefont{M.~J.} \bibnamefont{Gillan}},
  \bibnamefont{and} \bibinfo{author}{\bibfnamefont{G.~D.} \bibnamefont{Price}},
  \bibinfo{journal}{J. Chem. Phys.} \textbf{\bibinfo{volume}{118}},
  \bibinfo{pages}{10174} (\bibinfo{year}{2003}).

\bibitem[{\citenamefont{Shan et~al.}(2003)\citenamefont{Shan, Wang, Knoesel,
  Bonn, and Heinz}}]{SWK03}
\bibinfo{author}{\bibfnamefont{J.}~\bibnamefont{Shan}},
  \bibinfo{author}{\bibfnamefont{F.}~\bibnamefont{Wang}},
  \bibinfo{author}{\bibfnamefont{E.}~\bibnamefont{Knoesel}},
  \bibinfo{author}{\bibfnamefont{M.}~\bibnamefont{Bonn}}, \bibnamefont{and}
  \bibinfo{author}{\bibfnamefont{T.~F.} \bibnamefont{Heinz}},
  \bibinfo{journal}{Phys. Rev. Lett.} \textbf{\bibinfo{volume}{90}},
  \bibinfo{pages}{247401} (\bibinfo{year}{2003}).

\bibitem[{\citenamefont{Bersch et~al.}(2008)\citenamefont{Bersch, Rangan,
  Bartynski, Garfunkel, and Vescovo}}]{BRB08}
\bibinfo{author}{\bibfnamefont{E.}~\bibnamefont{Bersch}},
  \bibinfo{author}{\bibfnamefont{S.}~\bibnamefont{Rangan}},
  \bibinfo{author}{\bibfnamefont{R.~A.} \bibnamefont{Bartynski}},
  \bibinfo{author}{\bibfnamefont{E.}~\bibnamefont{Garfunkel}},
  \bibnamefont{and} \bibinfo{author}{\bibfnamefont{E.}~\bibnamefont{Vescovo}},
  \bibinfo{journal}{Phys. Rev. B} \textbf{\bibinfo{volume}{78}},
  \bibinfo{pages}{085114} (\bibinfo{year}{2008}).

\bibitem[{\citenamefont{Allen}(1992)}]{Allen92}
\bibinfo{author}{\bibfnamefont{J.~E.} \bibnamefont{Allen}},
  \bibinfo{journal}{Phys. Scripta} \textbf{\bibinfo{volume}{45}},
  \bibinfo{pages}{497} (\bibinfo{year}{1992}).

\bibitem[{\citenamefont{Lampe et~al.}(2003)\citenamefont{Lampe, Goswami,
  Sternovsky, Robertson, Gavrishchaka, Ganguli, and Joyce}}]{LGS03}
\bibinfo{author}{\bibfnamefont{M.}~\bibnamefont{Lampe}},
  \bibinfo{author}{\bibfnamefont{R.}~\bibnamefont{Goswami}},
  \bibinfo{author}{\bibfnamefont{Z.}~\bibnamefont{Sternovsky}},
  \bibinfo{author}{\bibfnamefont{S.}~\bibnamefont{Robertson}},
  \bibinfo{author}{\bibfnamefont{V.}~\bibnamefont{Gavrishchaka}},
  \bibinfo{author}{\bibfnamefont{G.}~\bibnamefont{Ganguli}}, \bibnamefont{and}
  \bibinfo{author}{\bibfnamefont{G.}~\bibnamefont{Joyce}},
  \bibinfo{journal}{Phys. Plasma} \textbf{\bibinfo{volume}{10}},
  \bibinfo{pages}{1500} (\bibinfo{year}{2003}).

\bibitem[{\citenamefont{Khrapak et~al.}(2005)\citenamefont{Khrapak, Ratynskaia,
  Zobnin, Usachev, Yaroshenko, Thoma, Kretschmer, Hoefner, Morfill, Petrov
  et~al.}}]{KRZ05}
\bibinfo{author}{\bibfnamefont{S.~A.} \bibnamefont{Khrapak}},
  \bibinfo{author}{\bibfnamefont{S.~V.} \bibnamefont{Ratynskaia}},
  \bibinfo{author}{\bibfnamefont{A.~V.} \bibnamefont{Zobnin}},
  \bibinfo{author}{\bibfnamefont{A.~D.} \bibnamefont{Usachev}},
  \bibinfo{author}{\bibfnamefont{V.~V.} \bibnamefont{Yaroshenko}},
  \bibinfo{author}{\bibfnamefont{M.~H.} \bibnamefont{Thoma}},
  \bibinfo{author}{\bibfnamefont{M.}~\bibnamefont{Kretschmer}},
  \bibinfo{author}{\bibfnamefont{H.}~\bibnamefont{Hoefner}},
  \bibinfo{author}{\bibfnamefont{G.~E.} \bibnamefont{Morfill}},
  \bibinfo{author}{\bibfnamefont{O.~F.} \bibnamefont{Petrov}},
  \bibnamefont{et~al.}, \bibinfo{journal}{Phys. Rev. E}
  \textbf{\bibinfo{volume}{72}}, \bibinfo{pages}{016406}
  (\bibinfo{year}{2005}).

\bibitem[{\citenamefont{Carstensen et~al.}(2011)\citenamefont{Carstensen, Jung,
  Greiner, and Piel}}]{CJG11}
\bibinfo{author}{\bibfnamefont{J.}~\bibnamefont{Carstensen}},
  \bibinfo{author}{\bibfnamefont{H.}~\bibnamefont{Jung}},
  \bibinfo{author}{\bibfnamefont{F.}~\bibnamefont{Greiner}}, \bibnamefont{and}
  \bibinfo{author}{\bibfnamefont{A.}~\bibnamefont{Piel}},
  \bibinfo{journal}{Phys. Plasmas} \textbf{\bibinfo{volume}{18}},
  \bibinfo{pages}{033701} (\bibinfo{year}{2011}).

\end{thebibliography}

\end{document}